\documentclass[preprint,journal]{vgtc}       

\ifpdf
  \pdfoutput=1\relax                   
  \usepackage{graphicx}                
  \DeclareGraphicsExtensions{.pdf,.png,.jpg,.jpeg} 
\else
  \usepackage{graphicx}                
  \DeclareGraphicsExtensions{.eps}     
\fi%

\graphicspath{{figures/}{pictures/}{images/}{./}} 

\usepackage{microtype}                 
\PassOptionsToPackage{warn}{textcomp}  
\usepackage{textcomp}                  
\usepackage{amsmath}
\usepackage{amsfonts}
\usepackage{amsthm}
\usepackage[ruled,linesnumbered]{algorithm2e}
\usepackage{times}   
\usepackage{xcolor}
\usepackage{enumitem}

\usepackage{cite}                      
\usepackage{tabu}                      
\usepackage{booktabs}                  
\usepackage{float}
\usepackage{inconsolata}

\usepackage{setspace}
\usepackage[hang,flushmargin]{footmisc} 

\definecolor{revisecolor}{HTML}{000000}  
\newcommand{\revise}[1]{\textcolor{revisecolor}{#1}}

\definecolor{TODOcolor}{HTML}{0000CC}

\onlineid{1096}

\vgtccategory{Research}
\vgtcpapertype{application/design study}

\title{A Visual Analytics Framework for Explaining and Diagnosing Transfer Learning Processes}

\author{Yuxin Ma, Arlen Fan, Jingrui He, Arun Reddy Nelakurthi, Ross Maciejewski}
\authorfooter{
\item
\vspace{-1.9mm}
Y. Ma, A. Fan and R. Maciejewski are with Arizona State University. E-mail: \{yuxinma,afan5,rmacieje\}@asu.edu.
\item
J. He is with the University of Illinois at Urbana-Champaign. E-mail: jingrui@illinois.edu.

\item
A. R. Nelakurthi is with Samsung Research America. E-mail: arunreddy.nelakurthi@gmail.com.
}

\abstract{
\revise{
Many statistical learning models hold an assumption that the training data and the future unlabeled data are drawn from the same distribution. However, this assumption is difficult to fulfill in real-world scenarios and creates barriers in reusing existing labels from similar application domains. Transfer Learning is intended to relax this assumption by modeling relationships between domains, and is often applied in deep learning applications to reduce the demand for labeled data and training time. Despite recent advances in exploring deep learning models with visual analytics tools, little work has explored the issue of explaining and diagnosing the knowledge transfer process between deep learning models. In this paper, we present a visual analytics framework for the multi-level exploration of the transfer learning processes when training deep neural networks. Our framework establishes a multi-aspect design to explain how the learned knowledge from the existing model is transferred into the new learning task when training deep neural networks. Based on a comprehensive requirement and task analysis, we employ descriptive visualization with performance measures and detailed inspections of model behaviors from the statistical, instance, feature, and model structure levels. We demonstrate our framework through two case studies on image classification by fine-tuning AlexNets to illustrate how analysts can utilize our framework.}

} 


\teaser{
    \centering
    \includegraphics[width=0.97\linewidth]{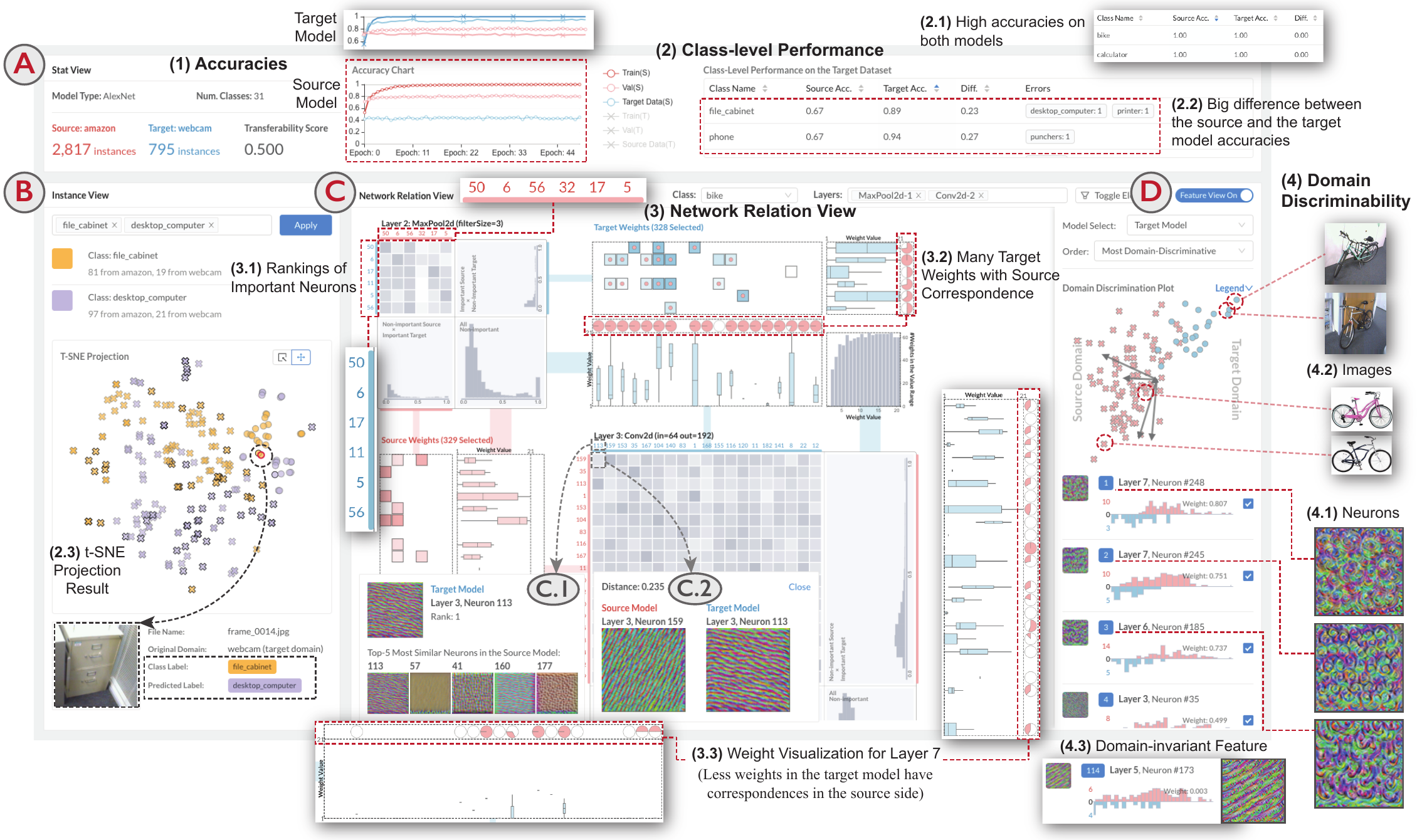}
    \caption{\revise{The} \revise{visual analytics for transfer learning interface consists of four components: (A) statistical information summary, (B) the instance view, (C) the network relation view, and (D) the feature view. For the Office-31 dataset, (1) the prediction accuracy of the target model on the source dataset is lower than the source model; (2) for the target dataset, classes such as \texttt{file\_cabinet} and \texttt{phone} have a large performance gap between the source and the target models; (3) the neuron similarity matrices and the weights are presented; (4) some neurons have high domain discriminability, while Neuron \#173 in Layer 5 is domain-invariant.}}
    \label{fig:teaser}
}

\vgtcinsertpkg

\begin{document}



\firstsection{Introduction}
\label{sec:introduction}

\maketitle

\begin{spacing}{1.0}


Machine learning approaches have achieved outstanding success in various fields including image recognition~\cite{krizhevsky2012imagenet}, natural language processing~\cite{Young2018nlp}, and question-answering systems~\cite{antol2015vqa}. However, in real-world applications, the collected labels used to train machine learning models may become quickly outdated, and collecting new labels can be cost prohibitive~\cite{pan2008transfer,Lu2018}. Furthermore, directly reusing expired or non-related labeled data from other domains may inject bias into the modeling process.
To reduce the cost of labeling new instances and minimize the threat of bias, researchers have proposed transfer learning, a meta-strategy to relax the assumption of independent and identical distribution (i.i.d) and mitigate data insufficiency~\cite{Pan2010a}. From a non-technical perspective, transfer learning is inspired by the knowledge transfer phenomenon in humans~\cite{Weiss2016} where, for example, an experienced car driver can learn to operate boats with less effort by mapping the control mechanism, or a pianist may master other musical instruments faster than new learners. Transfer learning is seen as a promising approach in the deep learning community, where the high demand for labels and long training times of models can be bottlenecks for model deployment~\cite{Tan2018}. By recycling labeled instances and pre-trained networks from related data domains, model accuracy can be improved and training time reduced as compared to training neural networks with new instances from scratch~\cite{Yosinski2014,Tan2018}.

While there have been successful applications of transfer learning in deep neural networks~\cite{krizhevsky2012imagenet,raghu2019transfusion}, understanding and interpreting the knowledge transfer processes is a critical step in the transfer learning process. Measuring the performance of the transferred model with conventional metrics is not enough to ensure that the knowledge transfer will be robust. Model developers need to understand \textit{what} knowledge has been transferred from the old model and \textit{how} the knowledge is reused in a particular form in the new model to boost its performance, which cannot be answered with simple statistical measures~\cite{tensorboard}. 


Given the recent success of visual analytics for explainable AI (XAI)~\cite{Bertini2009,Endert2017,Lu2017,Liu2017,Hohman2018}, we propose a visual analytics framework to explain and diagnose transfer learning processes in deep neural networks. \revise{The research challenges are summarized to characterize the exploration routines and essential knowledge users want to extract. Analytical tasks are derived in accordance with the requirements in a hierarchical manner and act as a guidance of the framework design.} While existing visual analytics tools focus on individual models~\cite{Kahng2017a,Wang2019deepvid,Liu2017a,Hohman2019a}, our approach addresses the task of revealing relationships between multiple models and utilizes a multi-faceted visualization scheme to present the transferred knowledge with respect to data, feature, and domain levels. \revise{Specifically, we propose an interactive interface consisting of two modules: descriptive visualization of statistical measures and detailed inspection of model behaviors. The statistical information of the models before and after transfer are revealed to provide an overall measure of how successful the transfer is, Figure~\ref{fig:teaser} (A). For detailed inspection components in Figure~\ref{fig:teaser} (B-D), the instance view depicts essential information on data distributions and predictions in the new model, the matrix-based network relation view is designed to reveal similarities and differences of the critical neural network components, and the feature view utilizes domain discriminability to support the exploration of shared knowledge hidden in the extracted features. To demonstrate our framework, we provide case studies and expert interviews in image classification tasks where a fine-tuning method is adopted on AlexNets.} Our contributions include:

\begin{itemize}[topsep=0.3em,partopsep=0em,parsep=0.2em,itemsep=0em,leftmargin=*,labelsep=0.4em]
    \item A visual analytics framework that supports the interpretation and diagnosis of the transfer learning processes in deep learning models;
    \item A suite of visualization designs that illustrate the transferred knowledge from the data, the model, and the feature levels.
\end{itemize}





    


\section{Related Work}
\label{sec:related_work}

In this section, we review related work on model interpretation and comparison as well as visual analytics for deep learning.

\vspace{0.8mm} \noindent \textbf{Model-Agnostic and Model-Specific Analysis.}\; A variety of work from the visual analytics community has focused on supporting model transparency for XAI. Previous work can be roughly classified into two main categories: model-agnostic design and model-specific design. Model-agnostic approaches consider the models to be black boxes where the internal learning processes are opaque to users. Here, the focus is on interpreting the models using the input data, output predictions, and performance measures, as such metrics are common across all classes of models. Manifold~\cite{Zhang2018}, ModelTracker~\cite{Amershi}, and Squares~\cite{Ren2016} utilize the predictions of labels from classifiers as well as performance measures to support model comparison, performance debugging, feature importance analysis, and instance-level explanations. FairSight~\cite{Ahn2019}, FairVis~\cite{Cabrera2019}, and the What-If Tool~\cite{Wexler2019} apply fairness metrics to evaluate whether algorithmic disparities of certain populations occur in the predictions. Prospector~\cite{Krause} enhances the model interpretability by demonstrating feature importance with partial dependence and localized inspection. 

Model specific designs focus on improving model transparency by revealing the inner workings of models in a white-box manner. For example, tree-based models are considered to be highly-interpretable models because of the natural presentation of decision criteria. For less-interpretable models, such as support vector machines~\cite{Ma2017} and artificial neural networks~\cite{Tzeng2005}, methods have been proposed to expose the core structures of the model (e.g., support vectors or neurons) or utilize interpretable surrogate models~\cite{Ming2018}. Compared with the model-agnostic approaches, the strategy of ``opening the black box''~\cite{Muhlbacher2014} benefits the advanced users by providing insights into underlying patterns hidden inside the models. However, the strong bindings between the internal structures of models and the specialized visualization designs limit the universality of such model specific visual analytics systems~\cite{Ren2016}.

\vspace{0.8mm} \noindent \textbf{Model Comparison.}\; In the predictive visual analytics pipeline~\cite{Lu2017,lu2017recent}, \revise{model comparison is essential for model selection, where the best result is selected from various predictions associated with multiple parameter settings~\cite{Lu2017,Li2020compare}}. Statistical charts, including line charts and scatterplots, have been applied to visualize model statistics and provide an intuitive comparison of different results~\cite{Lu2017,Gleicher2017}. Many visual analysis approaches provide a detailed comparative analysis of multiple predictions. Work by Spinner et al.~\cite{Spinner2019} summarizes typical tasks in XAI and proposes a framework for interactive machine learning. In the model quality control stage, the model comparison component is implemented using different ``explainers'' to support comparative explanation. Visual analysis approaches in ensemble learning\cite{Talbot2009,Liu2017b,Zhao2019}, AutoML\cite{Wang2019atmseer,Cashman2020}, and choropleth classification under uncertainty~\cite{Zhang2017,Huang2019} consider the comparison of individual ensemble members as the core analytical task. For topic modeling, Alexander and Gleicher~\cite{Alexander2016} explore the design space of task-oriented topic model comparison and derive comparison tasks based on their single-model counterparts. In data clustering, the DICON~\cite{Cao2011} system adopts a glyph-based visualization design to interpret, evaluate, and compare high-dimensional statistical information of clusters. Pilh\"{o}fer et al.~\cite{Pilh2012} presents an algorithm to compare clustering results from different clustering models. Along with prediction comparisons, the INFUSE system~\cite{Krause2014} addresses the need for comparative analysis in feature selection and provides a visual analytics system to guide the optimization of feature sets in classification tasks.

\vspace{0.8mm} \noindent \textbf{Visual Analytics in Deep Learning.}\; Given the current popularity of deep learning models, numerous visual analytics frameworks have also been proposed to improve the issues of low interpretability and transparency unique to the deep learning process. 
These visual analytics tools have been designed for deep learning experts and end users~\cite{Hohman2018} while tackling numerous real-world issues, such as model vulnerability~\cite{Goodfellow2015,Ma2020}, data security~\cite{abadi2016deep}, trust~\cite{lim2019trust}, and fairness\cite{Ahn2019,Cabrera2019}.
Much of the previous visual analytics work for deep learning focuses on analyzing Convolutional Neural Network (CNN) structures in image classification~\cite{Kahng2017a,Liu2018a,Wang2019deepvid,Alsallakh2017,Selvaraju2017,Chung2016,Hohman2019a,Liu,Wongsuphasawat2017,Rauber2016,Pezzotti2017} or Recurrent Neural Networks (RNN) and its variants in NLP tasks~\cite{Kwon2018,Strobelt2019,Ming2017,Strobelt2016a}. 
For unsupervised Generative Adversarial Networks (GAN), AEVis~\cite{Liu2018}, GAN Lab~\cite{Kahng}, and GANViz~\cite{Wang2018a} support the exploration of how adversarial examples from the generative networks pass through the discriminative ones and trigger erroneous behaviors in the neurons which dramatically change the final predictions. Wang et al.~\cite{Wang2018} propose DQNVis for the investigation of training dynamics and action outcomes in deep reinforcement learning models. 

Along with learning tasks, visualization for deep neural networks also focuses on prediction performance, network structures, and the behaviors of intermediate hidden layers~\cite{Hohman2018,Wang2019genealogy}. The performance diagnosis approaches for general black-box models can be applied to deep neural networks as well, and some work extends the black box methods by introducing instance checking components~\cite{Kahng2017a}. For model-specific inspection, network structures are illustrated using node-link diagrams~\cite{Wongsuphasawat2017} where nodes represents layers, or important neurons, and edges represent the weights or filters that connect the layers or neurons~\cite{Liu,Liu2018a,Hohman2019a}. Behaviors of hidden layers and neurons are sometimes embedded into the structure diagram, the activation map~\cite{Selvaraju2017}, the representative instances or features~\cite{Ming2017,Hohman2019a}, or the clusters of neurons~\cite{Liu}.


While, the visual analytics community has developed a variety of methods for model interpretability, performance diagnosis, and model comparison, none of the aforementioned approaches have addressed the explainability and diagnosis of knowledge transfer between deep learning models. \revise{Recent works by Zeng et al.~\cite{Zeng2017}, Murugesan et al.~\cite{Murugesan}, and Ma et al.~\cite{Ma2017transfer} are the closest in spirit to our work with respect to model comparison and knowledge transfer. However, our framework goes beyond simple comparisons between models and analysis of traditional transfer learning techniques on shallow learning models to further reveal the relationships of inheritance and knowledge reuse in deep neural networks.}




\section{Background on Transfer Learning}
\label{sec:background}

Before presenting our visual analytics framework, we first define the basic concepts of the transfer learning processes.

\vspace{0.8mm} \noindent \textbf{Domain and Task.}\; There are two core concepts in transfer learning: \textit{domain}, and \textit{task}~\cite{Pan2010a}.

\begin{itemize}[topsep=0.1em,partopsep=0em,parsep=0.2em,itemsep=0em,leftmargin=*,labelsep=0.4em]
    \item Given a set of data instances $X = \{\boldsymbol{x}_i | \boldsymbol{x}_i \in \mathcal{X}, i \in [1, N]\}$ in the feature space $\mathcal{X}$, a domain $\mathcal{D}$ is a combination of $\mathcal{X}$ and the marginal probability distribution $P(X)$. Take for example, a collection of news articles. We can consider the instances $\boldsymbol{x}$ to be the corresponding word count vectors of the articles, and the feature space $\mathcal{X}$ the vocabulary of all the articles. Due to variations of wording and expressions, the word count distribution is often unique for a specific news source, leading to different $P(X)$ across publications.
    
    \item The concept of task describes the supervision information and the model that fulfills a specific learning goal. Formally, a task $\mathcal{T}$ consists of two components: the label space $\mathcal{Y}$ which represents all possible labels for the instances, and a decision function $y = f(\boldsymbol{x})$ learned from the labeled instances $\{(y_i,\boldsymbol{x}_i) | y \in \mathcal{Y}, \boldsymbol{x} \in \mathcal{X}, i \in [1, N]\}$. In our news article example, the goal is to perform sentiment analysis, leading to a label space $\mathcal{Y}$ of two elements such that each label $y_i$ for $\boldsymbol{x}_i$ can be either $Positive$ or $Negative$. The classifier $f(\cdot)$ is generated by fitting the labeled instances and can be further applied to unlabeled instances to predict their sentiment.
\end{itemize}

\noindent In the conventional machine learning setting, the assumption should be held that unlabeled data comes from the same distribution as the training data and share the same set of class labels. Formally, by denoting the unlabeled instances as $X' = \{\boldsymbol{x}'_i | \boldsymbol{x}'_i \in \mathcal{X}', i = [1, M]\}$ and their potential labels in the set of $\mathcal{Y}'$, we have $\mathcal{X}' = \mathcal{X}$, $P(X') = P(X)$, and $\mathcal{Y'} = \mathcal{Y}$, which leads to $\mathcal{D} = \mathcal{D}'$ and $\mathcal{T} = \mathcal{T}'$. In the context of our sentiment analysis example, this requirement means that the unlabeled new articles should be collected from the same or similar websites with nearly identical patterns of word distributions and semantics, and the choices of class labels still remain either $Positive$ or $Negative$.

\vspace{0.8mm} \noindent \textbf{Transfer Learning.}\; In practice, the assumption of aligned domains and tasks for training and prediction stages may not hold. In our example, if the articles with sentiment labels are all retrieved from a political news channel, a sentiment classifier trained on these articles may generate biased predictions on sports news due to a potentially divergent vocabulary used to express positive and negative ideas, i.e., $P(X_{\text{politics}}) \neq P(X_{\text{sports}})$. There are other types of mismatches between the domains and tasks, such as varied numbers of unique words between news channels ($\mathcal{X}_{\text{politics}} \neq \mathcal{X}_{\text{sports}}$, namely, different feature spaces) or different concepts of classes ($\mathcal{Y}_{\text{politics}} \neq \mathcal{Y}_{\text{sports}}$).

The purpose of transfer learning algorithms is to handle the mismatching issue by learning a decision function $f(\cdot)'$ in the new domain $\mathcal{D}'$ and task $\mathcal{T}'$ with the knowledge learned from the existing $\mathcal{D}$ and $\mathcal{T}$, where $\mathcal{D}' \neq \mathcal{D}$ or $\mathcal{T}' \neq \mathcal{T}$. In the transfer learning terminology, $\mathcal{D}$ and $\mathcal{T}$ are identified as the \textit{source domain} $\mathcal{D}_s$ and the \textit{source task} $\mathcal{T}_s$, and the new domain and the task as the \textit{target domain / task} ($\mathcal{D}_t$ and $\mathcal{T}_t$), respectively. To facilitate the notations in the following sections, we further denote $X_{\mathcal{D}_s}$ and $X_{\mathcal{D}_t}$ as the \textit{source} and the \textit{target dataset}.

\vspace{0.8mm} \noindent \textbf{Transfer Learning in Deep Neural Networks.} Researchers and practitioners have identified two major issues that impact the  efficiency of applying exising deep learning models to new problems~\cite{Tan2018}, including the need of massive labeled instances for the new task and the prohibitive computational overhead. These two issues lend themselves well to the transfer learning paradigm. For the lack of labeled instances, the relaxation of being in the same domain enables the recycling of similar data sources via instance re-weighting. For models, various works~\cite{Yosinski2014,Tan2018} tend to reuse learned parameters from existing models or utilize special network layers that extract common features. \revise{In our framework, we focus on the widely-used fine-tuning technique, which applies the trained network layer parameters in existing models to new ones as an initialization. For the type of models, we demonstrate our framework in the scenario of image classification where CNN-based neural networks are adopted.}



\section{Design Overview}
\label{sec:overview}
Given the key features of transfer learning in deep neural networks, we have designed a visual analytics framework to explain and investigate knowledge transfer between neural network models. \revise{Here, we summarize the research challenges, and a set of analytical tasks are derived from the challenges and used to guide our framework design and development.}



\subsection{\revise{Research Challenges}}

\revise{To identify the research challenges in visual analysis of deep transfer learning process, we conducted a literature review on deep transfer learning~\cite{Weiss2016,Tan2018,Pan2010a,Long2018,raghu2019transfusion} and listed several requirements for explaining the knowledge transfer processes. The list was then refined through detailed discussions with our domain experts (who serve as co-authors). Furthermore, we organized the challenges in a structured manner based on the multi-level topology proposed by Brehmer et al.~\cite{Brehmer2013}}, and we identified two key challenges when analyzing the transfer processes.

\vspace{0.8mm} \noindent \textbf{C1: Analytical Complexity.}\; Inspired by the question of ``why'' a task is performed~\cite{Brehmer2013}, analytical complexity describes how complicated the analysis is. We establish three levels of operations. 

\vspace{0.8mm} \noindent \textit{\textbf{C1.1}: Uncover the learned knowledge in the models in two domains}. To help make the transfer learning process more transparent, analysts need to understand how the models are trained in different domains.
    
\vspace{0.8mm} \noindent \textit{\textbf{C1.2}: Explore the similarities and differences between the source and the target model}, i.e., ``searching'' the desired match/mismatch patterns. Interpreting the transfer learning processes requires analysts to not only explore individual models and the associated data domain, but also to perform an in-depth exploration of \revise{how similar the target model is to the source model.} For example, analysts may need to investigate whether the patterns in the CNN's layer weights in the source model are still expressed in the target model, or whether a specific unlabeled instance receives the same prediction in both models. By comparing the similarities and differences between the source and the target model, analysts can gain insights into the underlying transfer mechanism.
    
\vspace{0.8mm} \noindent \textit{\textbf{C1.3}: Discover instances in the source domain that carry common knowledge}. A key issue in transfer learning is to interpret the transferred knowledge from the source domain to the target domain. In instance-based transfer learning algorithms, source data instances carrying domain-invariant characteristics can be considered to be a form of externalization of the shared knowledge that supports instance-based interpretation. Revealing such instances can facilitate the understanding of the transferred knowledge by providing concrete examples.



\vspace{0.8mm} \noindent \textbf{C2: Data Granularity.}\; Another important factor is the hierarchy of data types involved in the transfer process, namely, the ``what'' in the task typology~\cite{Brehmer2013}. Usually, the categorization is based on data types. In the context of interpretable machine learning~\cite{Zhang2018,Zhao2019,Spinner2019,Hohman2019b}, the following data types are considered:


\vspace{0.5mm} \noindent \textit{\textbf{C2.1}: Statistical Descriptions and Measures}, including the distribution of data instances and measurement of model prediction performance;

\vspace{0.5mm} \noindent \textit{\textbf{C2.2}: Data Instances}, including characteristics of specific data instances and their corresponding predictions in the models and domains; 

\vspace{0.5mm} \noindent \textit{\textbf{C2.3}: Model Structures and Parameters}, including the prediction mechanisms and features extracted by the neural networks.

\subsection{Analytical Tasks}

We further distill the following tasks based on the \revise{research challenges} of analytical complexity and data granularity.

\vspace{0.8mm} \noindent \textbf{T1: Summarize the Model Performances.} Summarizing data distributions and performance metrics is a fundamental prerequisite to start the analysis. Analysts may be interested in a coarse-grained comparison between the source and the target model, such as:

\begin{itemize}[topsep=0.1em,partopsep=0em,parsep=0.2em,itemsep=0em,leftmargin=*,labelsep=0.4em]
    \item What are the differences between the overall model accuracies evolved with the number of epochs? \revise{(\textit{\textbf{C1.1, C1.2, C2.1}})}
    \item How do the models perform when making predictions for different classes? \revise{(\textit{\textbf{C1.1, C2.1}})}
\end{itemize}

\vspace{0.8mm} \noindent \textbf{T2: Revealing Classification Results in the Instance Level.}
\begin{itemize}[topsep=0.1em,partopsep=0em,parsep=0.2em,itemsep=0em,leftmargin=*,labelsep=0.4em]
    \item How are different classes separated after passing through the trained models (class-wise comparison)? \revise{(\textit{\textbf{C1.3, C2.2}})}
    \item Inside a class, how are the source and the target instances distributed? \revise{(\textit{\textbf{C1.3, C2.2}})}
\end{itemize}

\vspace{0.8mm} \noindent \textbf{T3: Detailed Inspection and Comparison of Underlying Model Behaviors.}
Regarding the differences between the source and the target models, it is essential to support the comparative analysis of hidden neural network components, including:
\begin{itemize}[topsep=0.1em,partopsep=0em,parsep=0.2em,itemsep=0em,leftmargin=*,labelsep=0.4em]
    \item How are the two models \revise{related} on the core components in the neural networks, such as the learned weights and activations upon a specific group of data instances? \revise{(\textit{\textbf{C1.1, C1.2, C2.3}})}
    \item With respect to the extracted features in the network layers, are these features able to distinguish the corresponding domains of the data instances? Do the features carry domain-invariant patterns? \revise{(\textit{\textbf{C1.2, C1.3, C2.3}})} 
\end{itemize}

\begin{figure}[t!]
	\centering	
	\includegraphics[width=0.99\columnwidth]{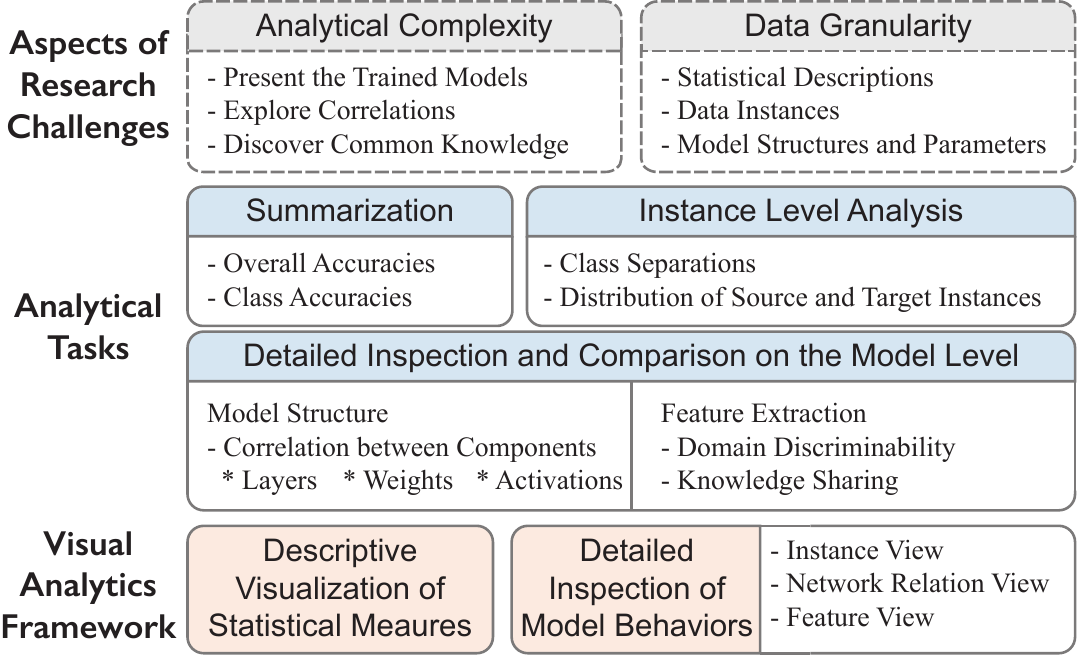}
	\caption{An overview of the \revise{research challenges}, tasks, and framework. The framework consists of two modules: descriptive visualization of statistical measures and detailed inspection of model behaviors.}
	\label{fig:overview_pipeline}
	\vspace{-6mm}
\end{figure}

\section{Visual Analytics Framework}
\label{sec:framework}

Based on the identified research challenges and the analytical tasks, we propose a visual analysis framework, Figure~\ref{fig:teaser}, for analyzing transfer learning processes from the source to the target domains. The analytical tasks are mapped to two modules in the framework:

\vspace{0.8mm} \noindent \textbf{Descriptive Visualization of Statistical Measures (T1).} This module acts as an entry point for the entire analysis pipeline where descriptions of model performances are visualized, Figure~\ref{fig:teaser} (A). By examining the prediction performances of the source and the target models on different classes, analysts can get an overview of the models and identify salient patterns in the measurements, such as classes with the highest or lowest prediction accuracy. These classes can be addressed in the details-on-demand exploration.

\vspace{0.8mm} \noindent \textbf{Detailed Inspection of Model Behaviors (T2, T3).} We have designed three views to support details-on-demand exploration:

\begin{enumerate}[topsep=0.08em,partopsep=0em,parsep=0.2em,itemsep=0em,leftmargin=*,labelsep=0.4em]
    \item The \textbf{Instance View} (Figure~\ref{fig:teaser} (B)) explains how the target model predicts data instances in selected classes (\textbf{T2}). A projection view is employed to show the separations among classes.
    \item The \textbf{Network Relation View} (Figure~\ref{fig:teaser} (C)) utilizes a matrix-based visualization design to reveal similarities of model components including filters and weights (\textbf{T3}).
    \item The \textbf{Feature View} (Figure~\ref{fig:teaser} (D)) visualizes the domain discriminability of feature extractors in the target model (\textbf{T3}). A feature discriminability plot is proposed to visualize how much domain-relevant information is carried by the filters, indicating the extent of knowledge sharing in the filter level.
\end{enumerate}

\noindent Figure~\ref{fig:teaser} shows the interface where analysts can freely explore the results and switch between views. The visual elements share the same color encoding scheme where the red color represents the data instances and models in the source domain and the blue color is the target domain.



\begin{figure*}[th]
	\centering	
	\includegraphics[width=2.00\columnwidth]{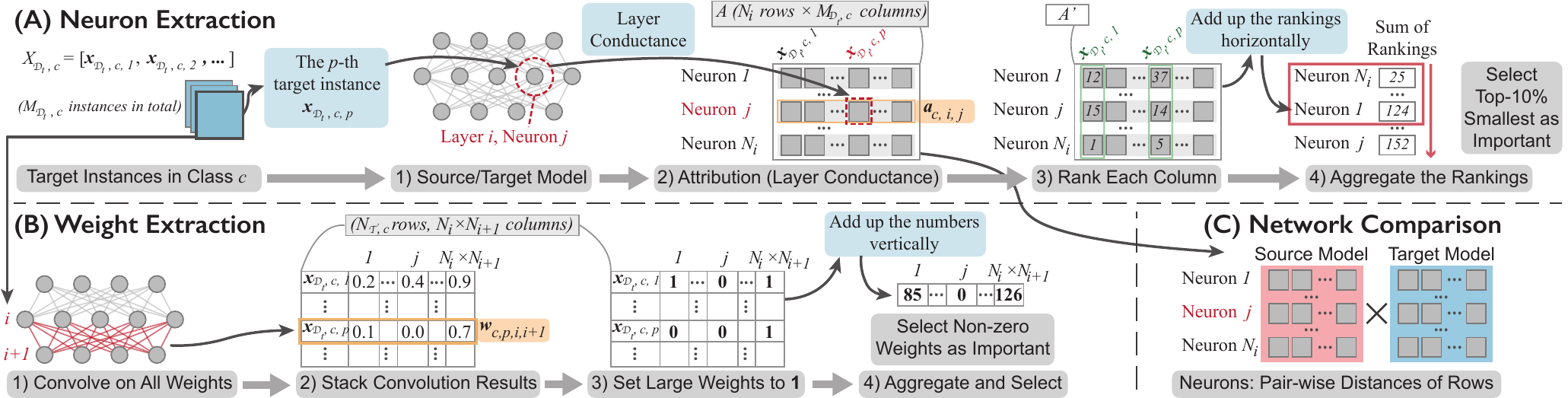}
	\caption{An illustration of the network abstraction and comparison procedures. (A) In each network layer, the important neurons are extracted based on the aggregated attribution (Layer Conductance) values on all the data instances. (B) Important weights are selected using a similar strategy. (C) Pairwise similarities between neurons in the same layers of the two models are computed.}
    \label{fig:network_extraction_precedures}
    \vspace{-5.8mm}
\end{figure*}

\subsection{Statistical Analysis of Model Performances}
The statistical measures provide an overview of how the source and the target models are trained and the predictions across all classes (\textbf{T1}). Additionally, the measures act as an entry point to reveal interesting clues and guide further detailed analysis.



\vspace{0.8mm} \noindent \textbf{\revise{Overall Prediction Accuracies and Transferability Score.}} We collect accuracy values in each epoch from both models on three different datasets: 1) the source training data, 2) the target training data, and 3) the validation set from the corresponding domain. Hence, each model derives three time series along the evolvement of the training epochs, which are then visualized as lines in a line chart. As shown in Figure~\ref{fig:teaser} (A), the horizontal axis indicates the number of epochs, while the accuracy values, ranging from zero to one, are mapped to the vertical axis. The cross and circle symbols on the lines represent the values from the source model and the target model, respectively. The line colors correspond to the dataset origin of the accuracy series. The suffixes of the legend entries indicate the corresponding models of the accuracy series. \revise{Along with the detailed illustration of the accuracy values, we also provide a transferability score on the left side of the line chart to give an overall quality measure of the transfer process. The score is defined as the difference between the best target and source model accuracies on the target dataset. A positive score indicates a performance boost because of the higher accuracy for the target model, and vice versa for negative scores.}

\vspace{0.8mm} \noindent \textbf{\revise{Confusion Table.}} To visualize the class-level performances, we compute the confusion matrix of the trained model on the target validation set and \revise{list measures in the confusion table}, Figure~\ref{fig:teaser} (A), comprised of five columns: 1) the class names; 2, 3) the accuracies of the two models on the target dataset; 4) the differences of the two accuracies on the same row, and; 5) the classes that the instances are misclassified into. Analysts can sort the rows based on the values in the column.

\subsection{Instance Analysis}

The instance view shows the data distributions of selected classes from the two domains and provides a detailed analysis of the relationships between instances with different attributes (\textbf{T2}). Instead of analyzing the original instances, i.e., raw pixels of the images, we focus on their embedding vectors extracted from the neural network layers. Thus, each instance is passed through the target model, and the activation vectors right before the fully-connected layers is used as the embedding vector. In our framework, we choose t-SNE~\cite{maaten2008visualizing} to visualize the network embeddings of the corresponding instances. 



In Figure~\ref{fig:teaser} (B), the instance view consists of three regions: a class selector, a projection scatterplot, and a detail view. Once the desired classes are selected, the details of the classes will be listed, and the projection result containing all the instances from the selected classes will be plotted. In the scatterplot, the colors of the glyphs indicate their class labels. To visualize the predictions made by the target model, the glyph borders for the mispredicted instances are set to dark gray. Similar to the encoding in the accuracy chart, we use crosses and circles for the instances from the source and the target domain, respectively. When hovering the mouse pointer on a glyph, the details of the corresponding instance are listed, including the original image, the domain of origin, the ground truth label, and the prediction made by the target model.

\subsection{Neural Network Component Analysis}
\label{sub:neural_network_component_analysis}

Since the learned knowledge is carried in the neural network parameters and outputs, including weights and activations, revealing the core components in the network layers can help analysts understand the critical patterns captured by the models in each domain and how a model predicts a specific instance (\textbf{T3}). Furthermore, comparing the components between the source and the target models can assist the analysis of the relationships between the two domains (\textbf{T3}). We propose a network abstraction and comparison method as well as the network relation view, Figure~\ref{fig:network_extraction_precedures}, to support the exploration of the neural networks and the \revise{similarities} between the two models.

\subsubsection{Network Abstraction}
One issue in visualizing the network layers is the excessive number of neurons and weights inside and between layers. In the network abstraction stage, we focus on extracting the essential structures that best represent the learned patterns and knowledge hidden in the networks. This stage consists of two steps: 1) the extraction of important neurons, and 2) the extraction of valuable weights.

\vspace{0.8mm} \noindent \textbf{Neuron Extraction} \revise{(Figure~\ref{fig:network_extraction_precedures} (A))}. The purpose of extracting important neurons is to rank the neurons in the same layer by importance. 

\noindent\revise{\textbf{Figure~\ref{fig:network_extraction_precedures} A - 1)}}\; \revise{For the target dataset $X_{\mathcal{D}_t}$, we denote the subset of data instances with the same class label $c$ as $\boldsymbol{x}_{{\mathcal{D}_t}, c, p}, p \in [1, M_{{\mathcal{D}_t}, c}]$, where $M_{{\mathcal{D}_t}, c}$ represents the number of such instances. For the $j$-th neuron in layer $i$ with $N_i$ neurons in total, we compute all the attribution values on all the $M_{{\mathcal{D}_t}, c}$ data instances in class $c$, i.e., $\{\boldsymbol{x}_{{\mathcal{D}_t}, c, p} | p \in [1, M_{{\mathcal{D}_t}, c}]\}$, resulting in an attribution vector $a_{c, i, j}$ for neuron $j$ in layer $i$ with the dimension of $N_{\mathcal{D}_t}$.} Usually, the attribution values are represented by using the activation values of the data instances on this neuron~\cite{Liu,Hohman2019a}. However, recent studies have identified limitations when directly applying activation values, especially on extracting important neurons~\cite{dhamdhere2018important,Leino2018}. To improve the ranking confidence, we use Layer Conductance~\cite{dhamdhere2018important,shrikumar2018computationally} as the attribution of neurons. Since the Layer Conductance result for each neuron in convolutional layers is a 2-D matrix, we use the maximum value in the matrix to simplify the representation of attributions.
    
\noindent\revise{\textbf{Figure~\ref{fig:network_extraction_precedures} A - 2)}}\; Then, we vertically stack the $a_{c, i, j}$ for all the $N_i$ neurons in to a matrix $A_{c, i}$ with a dimension of \revise{$N_i \times M_{{\mathcal{D}_t}, c}$}. Thus, each column represents the Layer Conductance attribution vector for an individual data instance from all neurons.
    
\noindent\revise{\textbf{Figure~\ref{fig:network_extraction_precedures} A - 3)}}\; Next, for each column, the neurons are ranked by the Layer Conductance values in ascending order, resulting in a ranking matrix $A'_{c, i}$. A neuron with a larger attribution value on a data instance receives a higher rank. Note that the tied values share an average ranking number. 

\noindent\revise{\textbf{Figure~\ref{fig:network_extraction_precedures} A - 4)}}\; Finally, the ranking matrix $A'_{c, i}$ is horizontally aggregated by summing up all the ranks in the same row, denoted as $\boldsymbol{r}'_{c, i}$. In this way, we use $\boldsymbol{r}'_{c, i}$ as a representation of neuron importance in layer $i$ on class $c$. To select a set of the most effective neurons, we select the neurons with the top-$k$ largest aggregated ranks. In our framework, $k$ is set to $10\%$ of the total number of neurons in the layer.

\vspace{0.8mm} \noindent \textbf{Weight Extraction} \revise{(Figure~\ref{fig:network_extraction_precedures} (B))}. Along with extracting the important neurons for each class, we also identify the links between consecutive layers which are frequently active. For layer $i$ with $N_i$ neurons and layer $i+1$ with $N_{i+1}$ neurons, there are $N_i \times N_{i+1}$ pair-wise weights between the two groups of neurons. We employ a strategy similar to the neuron selection procedure for selecting important weights:

\noindent\revise{\textbf{Figure~\ref{fig:network_extraction_precedures} B - 1)}}\; First, for each target instance $\boldsymbol{x}_{{\mathcal{D}_t}, c, p}$ in the class $c$, we compute the weight values by convolving the weight kernels on the activation values from layer $i$, resulting in an $N_i \times N_{i+1}$-dimensional vector, $\boldsymbol{w}_{c, p, i, i+1}$, that represents the activated levels on each weight.

\noindent\revise{\textbf{Figure~\ref{fig:network_extraction_precedures} B - 2)}}\; Under the vertical stacking scheme, we get a matrix containing \revise{$M_{{\mathcal{D}_t}, c}$} rows and $N_i \times N_{i+1}$ columns. Since the number of weights is relatively large, we apply a similar weight selection scheme as described in Hohman et al.~\cite{Hohman2019a} to reduce the computational cost. For each row, we set the cells with the top-$k$ weight values in the row to 1, and other cells to 0. In this way, we filter out the weights that are not important in classifying instances in class $c$. The matrix is further aggregated by adding up the replaced values vertically, bringing an importance vector $\boldsymbol{w}_{c, i, i+1}$ with $N_i \times N_{i+1}$ values which record the counts of being important for each weight.

\noindent\revise{\textbf{Figure~\ref{fig:network_extraction_precedures} B - 3)}}\; Finally, the weights with the top-$k$ importance values in $\boldsymbol{w}_{c, i, i+1}$ are used as the representative links between layer $i$ and $i+1$.
    

\subsubsection{Network Comparison}

By employing the neuron and weight extraction procedures to the source and the target models respectively, two sets of important neurons and weights can be extracted. To tackle the task of model comparison in T3, this step focuses on how to reveal the similarities and differences between the network components from the two models.


With respect to the neurons, we audit how similar the behaviors of the neurons are for the neurons in the two corresponding layers, \revise{Figure~\ref{fig:network_extraction_precedures} (C)}. In the neuron extraction step, we have described how to retrieve the attributions for each neuron on all the target data instances, namely, $A_{c, i} = [a_{c, i, j}], j \in N_i$. We use $A_{\mathcal{D}_s, c, i}$ and $A_{\mathcal{D}_t, c, i}$ to represent the matrices from the source and the target models, respectively. We argue that two neurons from the two domains are considered similar once they have similar distributions of attribution values across all the target instances. Here, we use the cosine distances to measure the pair-wise similarities of the neurons in $A_{\mathcal{D}_s, c, i}$ and $A_{\mathcal{D}_t, c, i}$. The resulting similarity matrix $S_{c, i}$ records how close the attribution behaviors are between the neurons from different models.

\begin{figure}[tbh]
	\centering	
	\includegraphics[width=0.99\columnwidth]{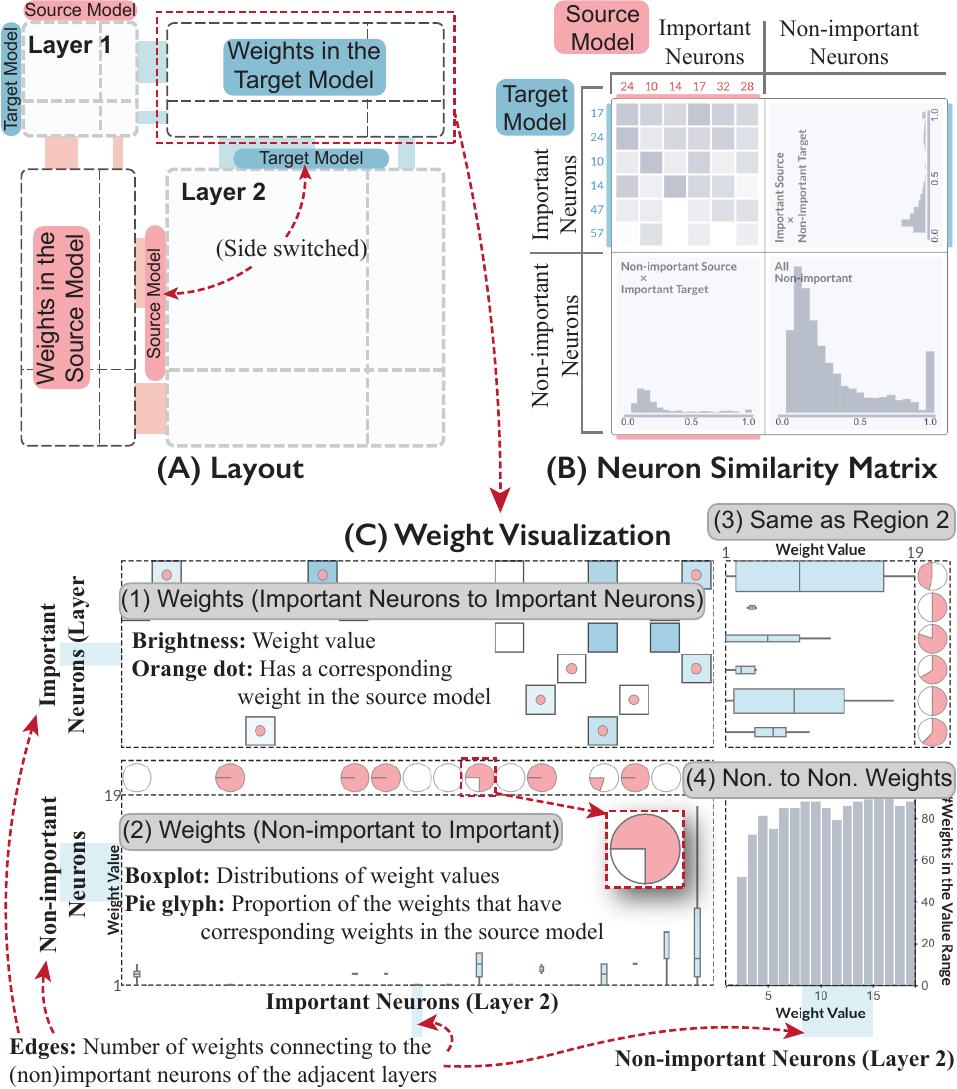}
	\caption{\revise{Visual} \revise{design and layout of the neuron similarity matrices and the weights. (A) The matrices are placed in a zig-zag manner to allow the weight components to be placed in the adjacent corners between the two consecutive matrices. (B) The rows and columns of the similarity matrix are grouped into important and non-important neurons, resulting in four regions. (C) A similar grouping scheme is applied to the weights.}}
    \label{fig:network_relation_view}
    \vspace{-5mm}
\end{figure}

The similarities of weights are built upon the neuron similarity results. For each selected important weight $w_{\mathcal{D}_t, i, i+1}$ in the target model that connects the neuron $n_1$ in layer $i$ and $n_2$ in layer $i+1$, we first find the corresponding most similar neurons of $n_1$ and $n_2$ (denoted as $n'_1$ and $n'_2$) in the same layers in the source model by looking up the matrices $S_{c, i}$ and $S_{c, i+1}$. Then, we see if the weight that connects $n'_1$ and $n'_2$ is in the important weight list in the source model. In this way, we can find whether an important weight in the target model is inherited from the source side and is valuable in making predictions.

\subsubsection{Visualization and Interactions}

The network relation view is designed to visualize the extracted neurons and weights between the source and the target models. 

\vspace{0.8mm} \noindent \textbf{Neuron Similarity Matrix.} This view leverages a diagonal layout where the similarity matrices are placed from the top left corner to the bottom right corner along the diagonal line, Figure~\ref{fig:network_relation_view} (A). In each matrix, the similarity values are linearly scaled to the cell brightness, where zero is represented by the highest brightness and one the lowest. In our initial design, all the neurons are listed in the matrix without considering their importance. However, such a matrix may be too large to be displayed in the canvas. We improved the scalability by grouping the rows and columns based on their importance in the neuron extraction step, dividing the matrix into four regions, Figure~\ref{fig:network_relation_view} (B). The top-left region shows the important neurons from both the source and the target models in the matrix form, while the other three regions are summarized into histograms of similarity value distributions. On the left and the top side of the matrices, the corresponding neuron indices are displayed with colors indicating the domain origins.

\vspace{0.8mm} \noindent \textbf{Weight Visualization.} In Figure~\ref{fig:network_relation_view} (A), the order of the domain origins in two consecutive matrices (Layer 1 and 2) are switched alternatively, leaving the space of weights in the source and the target model on the corresponding adjacent corners. Following the same splitting strategy with the matrices, the weights are also summarized into four regions, Figure~\ref{fig:network_relation_view} (C). In Region (1), the weights joining important neurons between both layers are rendered as an adjacency matrix. The rows and columns of the small squares are aligned with the corresponding important neurons in the similarity matrices. The brightnesses of the cells are mapped to the weight values, and the inner red dot indicates the existance of the corresponding important weight in the source model. Region (2) and (3) summarize the weight values into boxplots by aggregating the weights on the side of the important neurons. For example, the left side of Region (2) is next to the non-important neurons in Layer 1, while the bottom side connects to the important neurons in Layer 2. Thus, the vertical boxes show the weight value distributions connected with the same important neuron in Layer 2. The thicknesses of the rectangles in the boxes represent the number of weights. Pie chart glyphs are attached next to the boxes to show the proportions of weights with correspondence in the source model. For the weights connecting non-important neurons on both sides, we use a histogram in Region (4) to summarize the distribution of the weight values. Note that on the source model side (e.g., the bottom left corner in Figure~\ref{fig:network_relation_view} (A)), there are no red dots in the squares or pie chart glyphs since only the correspondences from target network components are considered.

\vspace{0.8mm} \noindent \revise{\textbf{Alternative Design.} We also considered utilizing node-link diagrams where the neurons in different layers are encoded as columns of nodes with weight links between consecutive columns~\cite{Liu,Hohman2019a}. However, isolating the source and the target models into two diagrams weakens the direct comparison of the corresponding layers and weights in the two neural networks. Furthermore, visual clutter can be easily generated when large numbers of neurons and weights exist in the layers, whereas the matrix layout can increase readability by providing a non-overlapped representation of nodes and links~\cite{VonLandesberger2011}.}

\vspace{0.8mm} \noindent \textbf{Interactions.} The network relation view supports various interactions on the similarity matrices and the weights. Users can change the desired class and network layers in the title bar. Clicking on the indices will open a detail panel, Figure~\ref{fig:teaser} (C.1), which shows a feature visualization~\cite{olah2017feature} of the neuron and the top-5 most similar neurons in the other model. By clicking on a square, the details of the associated weight are shown in a pop-up panel including the indices of the connected neurons as well as the weight feature map, Figure~\ref{fig:teaser} (C.2). The visual elements, including cells, boxplots, and pie glyphs, also provide pop-up details. \revise{To simplify the displayed elements, two switches are employed to toggle the appearance of non-important regions in the similarity matrices and the source weights.}



\subsection{Feature Analysis}
Deep neural networks are always considered as feature extractors that build new features from the original input space. As such, analysts often need to inspect what features are reconstructed in the hidden layers and how they are used in the classification process (\textbf{T3}). We have designed the feature analysis module to present the feature-related information and support the exploration of informative features.

\subsubsection{Domain Discriminability}
Usually, how a feature can discriminate the instances from different domains plays an essential role in a successful transfer. The feature extractors should be carefully investigated to see if they can reconstruct important common patterns shared in both domains. Most of the feature-based transfer learning techniques seek to build new feature transformations upon the source domain to create such shared features.

\setlength\textfloatsep{1.8mm}
\begin{algorithm}[htp]
    \label{alg:a-distance}
    \SetAlgoLined
    \KwData{
        $P$ selected neurons, $\{n_1, n_2, ..., n_{P}\}$; the source dataset, $X_{{\mathcal{D}_s}} = \{(\boldsymbol{x}_{{{\mathcal{D}_s}}, i}, y_{{{\mathcal{D}_s}}, i}) | i \in [1, N_{{{\mathcal{D}_s}}}] \}$; the target dataset $X_{\mathcal{D}_t} = \{(\boldsymbol{x}_{{\mathcal{D}_t}, i}, y_{{\mathcal{D}_t}, i}) | i \in [1, N_{\mathcal{D}_t}] \}$ 
    }
    
    \KwResult{The domain discriminability values for the $P$ neurons, $\{u_1, u_2, ..., u_P\}$}
    
    $A \leftarrow [\textrm{empty matrix}]_{(N_{\mathcal{D}_s} + N_{\mathcal{D}_t}) \times P}$ \tcp{Attribution values}

    $\boldsymbol{l} \leftarrow [\textrm{empty vector}]$ \tcp{Domain labels}

    \For{$i = 1:N_{\mathcal{D}_s}$}{ \label{alg-source-start}
        \For{$j = 1:P$}{
            $\boldsymbol{A}_{i, j} \leftarrow \textrm{LayerConductance}(n_j, \boldsymbol{x}_{{\mathcal{D}_s}, i})$
        }\label{alg-source-end}
        $\boldsymbol{l}_i \leftarrow 0$ \tcp{Label 0 for source instances} \label{alg-domain-label-1}
    }

    \For{$i = 1:N_{\mathcal{D}_t}$}{ \label{alg-target-start}
        \For{$j = 1:P$}{
            $\boldsymbol{A}_{i + N_{{\mathcal{D}_s}}, j} \leftarrow \textrm{LayerConductance}(n_j, \boldsymbol{x}_{{\mathcal{D}_t}, i})$
        }\label{alg-target-end}
        $\boldsymbol{l}_{i + N_{{\mathcal{D}_s}}} \leftarrow 1$ \tcp{Label 1 for target instances} \label{alg-domain-label-2}
    }

    $C = \textrm{LinearSVM}(A, \boldsymbol{l}, \textrm{10-fold cross validation})$ \label{alg-classifier}


    $\boldsymbol{u} \leftarrow [\textrm{feature weights of}$ $C]$ \tcp{Use the weights from the classifier as domain discriminability values}

    \caption{Computing the domain discriminability values.}

\end{algorithm}

To support the diagnosis of the transferred knowledge, we propose \textit{domain discriminability} to measure whether a learned feature on the neuron is domain-invariant or domain-dependent. This measure is inspired by the widely-used $\mathcal{A}$-distance~\cite{ben2007analysis} which estimates the differences of two data distributions with a linear classifier between the two groups of data. The computation of the domain discriminability for neurons is described in Algorithm~\ref{alg:a-distance}. First, we compute the Layer Conductance values for each selected neuron on the source and the target datasets, respectively (line~\ref{alg-source-start}-\ref{alg-source-end} and~\ref{alg-target-start}-\ref{alg-target-end}). In the matrix $A$, each row represents the attribution values for an instance on the selected neurons. Then, a domain label list $\boldsymbol{l}$ is prepared to specify the domain origins of each row (line~\ref{alg-domain-label-1} and~\ref{alg-domain-label-2}). Finally, a linear classifier $C$ is trained on the stacked attribution matrix and the domain label list (line~\ref{alg-classifier}). Acting as a feature selection method, the coefficients from the trained $C$ present how important a column in $A$, i.e., a selected neuron, can be in discriminating the domain origins of the data instances.

One potential limitation in computing the domain discriminability values is that the total number of neurons in the entire target neural network may be too large to train the domain classifier in an acceptable time. To reduce the computational cost, only the ``important neurons'' in each layer are selected as the input in Algorithm~\ref{alg:a-distance} since transfer methods often focus on the neurons with higher predictive power.



\subsubsection{Visualization and Interactions}

The design of the feature view, Figure~\ref{fig:teaser} (D), consists of two components: the feature ranking list, and the domain discriminability plot.

\vspace{0.8mm} \noindent \textbf{Feature Ranking List.} The details are listed for the neurons involved in the computation of domain discriminability, including the indices and layers of the neurons, the feature visualization of the neuron, and a histogram of Layer Conductance values for all the instances in the source and the target datasets. To differentiate the instances from two domains, the distribution of source instances are placed above the horizontal axis, while the target distribution is at the bottom. The colors of the bars map to the corresponding domains as well. Analysts can select whether to sort the neurons based on their domain discriminability values in ascending or descending order.

\vspace{0.8mm} \noindent \textbf{Domain Discriminability Plot.} To provide a detailed view of how the domains are discriminated by the neurons, we use a linear projection method to show the decision boundary of the domain classifier $C$ in Algorithm~\ref{alg:a-distance}. First, a linear projection matrix \revise{$W_{P \times 2}$} is constructed as $[\overrightarrow{\boldsymbol{u}}, \overrightarrow{\boldsymbol{
g}}] \in \mathbb{R}_{P \times 2}$, where \revise{$\boldsymbol{u}$} in the first column is the feature weights of $C$, and \revise{$\boldsymbol{g}$} the first principal component on $A$. By projecting $A$ onto a 2-D scatterplot with $A \cdot 
W$, the decision boundary of $C$ can be displayed on the horizontal direction, and the underlying patterns along the decision boundary can be revealed on the vertical axis. The domain origins for the rows in $A$ are mapped to the shapes and colors where red cross represents the source instances and blue circles the target ones. To further present the importance of each neuron, we draw the original axes in a biplot-like manner in the scatterplot. The lengths of the axes lines on the horizontal direction indicate their domain discriminability. We note that this may cause visual clutter of the axes lines when $P$, the number of neurons, is large. Thus, we only activate the top five neurons in the feature ranking list depending on the order selection, i.e., only the top five rows in \revise{$W$} and the corresponding columns in $A$ are considered. Additional neurons can be activated by clicking on the checkboxes in the corresponding rows of the feature ranking list.

\section{\revise{Case Study and Expert Interview}}
\label{sec:evaluation}

This section describes how our framework facilitates the understanding and exploration of transfer learning processes through applications in real-world datasets \revise{and feedback from domain experts. We implemented our framework with PyTorch for the deep learning library and React for the front-end framework. The data is transmitted in JSON format with RESTful APIs implemented with Flask.}


\begin{figure}[t]
	\centering	
    \includegraphics[width=1.00\columnwidth]{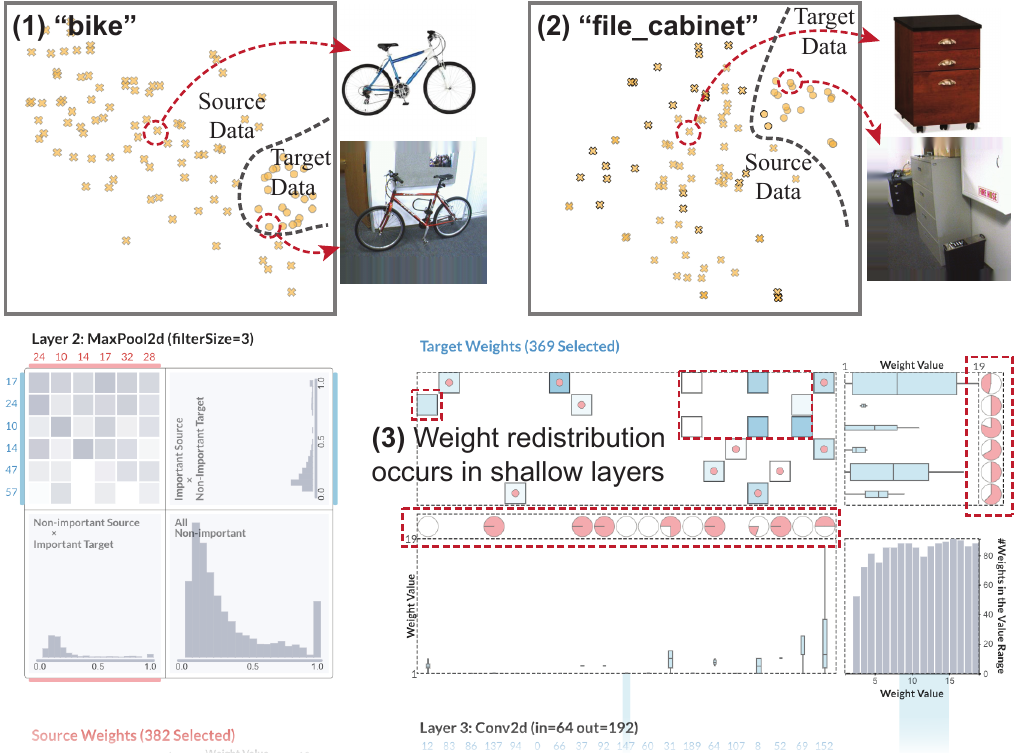}
	\caption{The t-SNE projection results for (1) \texttt{bike} and (2) \texttt{file\_cabinet} are presented. (3) Many of the target weights between Layer 2 and 3 have no associated important source weights, suggesting a weight redistribution in the target model.}
    \label{fig:teaser_sub1}
    \vspace{1.2mm}
\end{figure}

\subsection{Object Classification}
\label{sub:case1}

Office-31~\cite{saenko2010adapting} is a widely-used real-world dataset for demonstrating transfer learning algorithms. It contains 31 categories of images crawled from shopping websites and photo collections. In our experiment, we use the images from Amazon product pages (\textit{``amazon''}, 2817 images in total) as the source domain data, and photos from web cameras (\textit{``webcam''}, 795 images in total) as the target domain. For each dataset, we split the data into training and validation sets with the ratio of $85\%:15\%$, respectively. AlexNet~\cite{krizhevsky2012imagenet} was used as the backbone architecture for modeling, and it consists of five convolutional layers with the filter numbers of 96, 128, 384, 256, and 256, respectively. After AlexNet was trained on the \textit{amazon} dataset, we adapt the model to the \textit{webcam} data by using the trained parameters from the source domain AlexNet as an initialization.

\vspace{0.8mm} \noindent \textbf{Analysis of the Model Statistics (T1).} After the two models and datasets are loaded into our system, the brief summary of the model performances is depicted in Figure~\ref{fig:teaser}. We first check the accuracies on different datasets for the models. In the accuracy chart, Figure~\ref{fig:teaser} (1), we see that the source model performs well on the source dataset with high training and validation accuracies \revise{($\approx 0.98$ and $0.8$, respectively)}. However, the accuracy of the source model on the target dataset is $\approx 0.4$, indicating a big difference in patterns between the \textit{amazon} and the \textit{webcam} dataset. By checking the accuracies of the target model, we discover that the performance on its own domain (\textit{webcam}) is significantly higher than the source model, but it has a slightly worse accuracy on \textit{amazon} than the source model. The performance drop may be due to the loss of some unique pattern extractors in the target model for the source domain, which are replaced by the new knowledge specifically for the target data.

\begin{figure*}[t]
	\centering	
    \includegraphics[width=2.0\columnwidth]{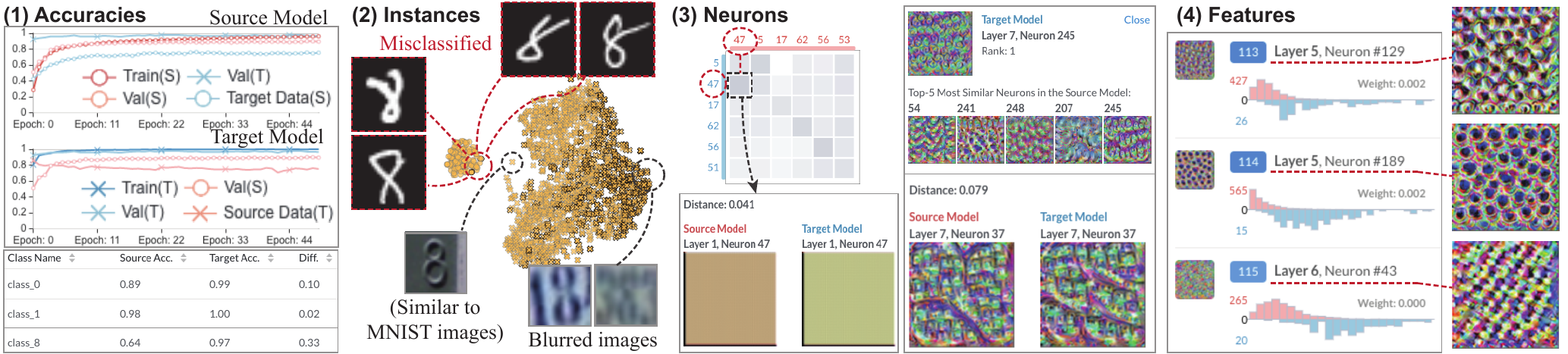}
	\caption{The result of the digit recognition datasets. (1) The trends of the accuracies are similar to the result in the first case study. (2) For the target model, the misclassified instances in SVHN and MNIST datasets are presented. (3) Several neurons show changes of the learned patterns in the target model. (4) The most domain-invariant neurons (features) indicate that the circular patterns are shared between the two domains.}
    \label{fig:case2}
    \vspace{-4mm}
\end{figure*}

\vspace{0.8mm} \noindent \textbf{Instance-Level Inspection (T2).} Apart from the overall accuracy analysis, we want to understand the prediction performance on the class and the instance levels. Here, we sort the third column of the class table to find the classes with the worst prediction accuracies by the target model, Figure~\ref{fig:teaser} (2). We find that some classes receive the same high accuracies from both models, such as \texttt{bike} and \texttt{calculator} (2.1), while classes like \texttt{file\_cabinet} and \texttt{phone} have the most diverse performances between the two models (2.2). We further examine their data distributions by activating \texttt{bike} and \texttt{file\_cabinet} in the instance view, respectively. In Figure~\ref{fig:teaser_sub1} (1), we observe that in the classes with similar performance on the two models, the images from the two domains share similar characteristics including visual angles and the appearance of the objects. However, the patterns significantly vary between the two domains, Figure~\ref{fig:teaser_sub1} (2).

Besides inspecting the interplay between classes, we rank the class table by target accuracies. Although most of the classes show nearly perfect prediction performances, several classes still contain mis-classified instances including \texttt{file\_cabinet} and \texttt{phone}. We select \texttt{file\_cabinet} and \texttt{desktop\_computer} in the instance view to show the erroneous predictions in \texttt{file cabinet}, Figure~\ref{fig:teaser} (2.3). In the t-SNE projection result, the distributions of the two classes are overlapped without a clear boundary, indicating a high chance of mis-classification.

\vspace{0.8mm} \noindent \textbf{In-depth Exploration of the Models (T3).} To reveal how the knowledge is shared in the classes with the same high accuracy or the diverse accuracy values, we further investigate the class \texttt{bike} and \texttt{file\_cabinet} in the network relation view and the feature view. After \texttt{bike} is selected in the network relation view (Figure~\ref{fig:teaser} (3)), we check the neuron similarity matrices from the shallow layers to the deep ones. In the shallower layers, such as Layer 2, Figure~\ref{fig:teaser} (3.1), the indices of the important neurons from the target model are similar to the source side, such as neuron 50, 6, and 17. Additionally, many of the source neurons with the same indices are also the ones most similar to their target counterparts, indicating no major changes in the functions of the neurons in the target model. This phenomenon still exists in deeper layers such as Layer 7, where the first five neurons from the two models are the same. However, the distributions of important weights vary drastically when the layer depth increases. By observing the cells with central red dots and the pie glyphs between Layer 2 and 3, Figure~\ref{fig:teaser} (3.2), we see that most of the weights in the target model have corresponding important weights in the source model. This suggests that most of the patterns hidden in the network links are reused, which also matches the findings of Yosinski et al.~\cite{Yosinski2014} that the shallow features are often reused in fine-tuning. However, the weight visualization shows a considerable change in the weights between Layer 6 and 7 where the proportions of the red arcs in most of the pie glyphs are much lower, Figure~\ref{fig:teaser} (3.3). We consider that although the neurons have similar rankings between both models, the weights are re-distributed between deeper layers in the target model in order to fit the new patterns in the target dataset. By inspecting the weights between Layer 2 and 3 in the class \texttt{file\_cabinet}, Figure~\ref{fig:teaser_sub1} (3), we find that the weight redistribution even occurs in the shallow layers. This indicates that the learned patterns diversify earlier between the source and the target model, resulting in the performance gap on the target dataset between the two models.

Finally, we explore the domain discriminability to find the common or unique features in the two domains. In the ranking result for \texttt{bike} ordered by the most domain-discriminative neurons, the top three features (Figure~\ref{fig:teaser} (4.1)) are used to extract flat ellipses, which may be related to the tires. By further checking the instances in the domain discrimination plot, Figure~\ref{fig:teaser} (4.2), we observe that most of the bike images in the source domain (\textit{amazon}, red crosses) are side-looking product profile images, while in the target domain (\textit{webcam}, blue circles) the bikes are recorded at various camera angles, causing different shapes of tires in the photos. This indicates that the tire shape patterns from the source domain are not fully transferred to the target model. Meanwhile, some features with low discriminability ranks depict patterns of handlebars and frames (e.g., Layer 5, Neuron 173, Figure~\ref{fig:teaser} (4.3)), which may be considered transferable between domains since these bike components exist in almost all photos in both domains.

\subsection{Digit Recognition}
\label{sub:case2}

In the second case study, we explore the knowledge transfer process between digit recognition datasets. We apply the Google Street View House Numbers dataset (SVHN) as the source domain, which contains natural scene photos of printed number labels. For the target domain, the MNIST dataset is used where the digits are in handwritten style and stored as grayscale images. We sampled 1500 instances per class in SVHN and 150 in MNIST, resulting in a source dataset with 15000 samples and a target dataset with 1500 samples in total. \revise{The purpose of downsampling is to mitigate scalability issues in computation and visualization, and these limitations are discussed in Section~\ref{sec:discussion}.} We use the same AlexNet architecture for both models.

\vspace{0.8mm} \noindent \textbf{Analysis of the Model Statistics (T1).} After the results are loaded, the accuracy chart and the class table are shown in Figure~\ref{fig:case2} (1). Similar to the lines in the first case, we find that the prediction accuracies of the two models show a diverse trend where each model performs well on its own domain but relatively poorly on the other domain. For the target model, this indicates that it has omitted some knowledge inherited from the source model and created new concepts specifically for the target dataset. In the class table, we find that some classes have a high accuracy in both models including \texttt{1} and \texttt{0}. However, there is a big difference between the two accuracy values from the source and the target model as well as several misclassified instances in the class \texttt{8}.

\vspace{0.8mm} \noindent \textbf{Instance-Level Inspection (T2).} Based on the findings in the class table, we want to examine the class of \texttt{8} in the instance view for more detailed explanations. In the t-SNE projection result for all the instances in SVHN and MNIST, Figure~\ref{fig:case2} (2), we observe that most of the mispredicted source instances (crosses with bold borders) are heavily blurred, which can potentially be a domain discrimination factor since the strokes in the handwritten digit images are much crisper. As for the misclassified instances, we observe that some irregular patterns presented in these images have prevented them from being recognized correctly, such as extra-long strokes, the unnecessarily big opening on the top, and the cropped bottom circles.

\vspace{0.8mm} \noindent \textbf{In-depth Exploration of the Models (T3).} We perform a detailed diagnosis of \texttt{8} in the network relation view and the feature view. After checking the feature visualization of the target model neurons in the similarity matrix of Layer 1, Figure~\ref{fig:case2} (3), we find that the extracted colors are slightly different between the target neurons and their most similar ones in the source model. Such differences become more significant in deeper layers where the corresponding neurons with the same indices in both models are not exclusively the most similar, such as the neuron 245 in Layer 7 whose most similar neuron in the source model is 54 instead of 245. Additionally, some neurons rank higher in the target model because of their unique characteristics to the MNIST dataset. For example, neuron 37 in the 5th place is used for extracting long free-style curves, which is common in the free handwritten images. However, its most similar counterpart, neuron 37 in the source model, receives a much lower ranking (rank 16). This suggests that the target model has adjusted the weights of several neurons to fit the new patterns emerging in the target data. In the feature view, Figure~\ref{fig:case2} (4), it can be observed that the neurons for extracting circular shapes are the most domain-invariant features, indicating that in the class \texttt{8}, the feature extractors for circles in the source model are reused in the target model.

\subsection{Expert Interviews}

\revise{Our framework was further evaluated by two machine learning practitioners, E1 and E2, whose expertise is in deep learning for computer vision and medical image analysis. In each interview, we first introduced the background, the tasks, and the interface of our framework. The analytical workflow was then explained with the two datasets in the case study. The experts were allowed to explore the datasets freely during the interview process. Finally, we collected comments on the analytical workflow and the visual interface. The two interviews lasted approximately 1.5 and 1 hour, respectively.}

\revise{The two experts agreed on the effectiveness of the workflow on analyzing transfer learning processes. They noted that it is an interesting aspect to investigate how the patterns learned in the source model are reused in the target model in an interpretable way. E1 commented that the multi-aspect analysis differs from conventional evaluation methods in deep learning where only statistical measures and data embeddings are considered. By examining the instances, network structures, and features in the coordinated views, analysts can gain insights into how the model parameters are reused in the target model and whether the learned patterns differ or were inherited from those in the source model. E2 pointed out that this framework can enable the diagnosis of per-class performance analysis and discover dominant features shared between the domains, which may further guide what types of existing labeled data can be reused in the current classification task.}

\revise{Our framework received positive feedback on the visual representation and interactions in the views. The experts observed that the accuracy chart and the confusion table act as a proper entry point for the experts. E2 mentioned that ``Checking accuracy values on the test datasets are common routines in our daily workflow. The per-class entries in the confusion table give me direct feedback on which classes receive bad performance, and I can continue investigating what is wrong in the detailed views.'' E1 addressed the usefulness of exploring model structures in the network relation view and the interplay between the detailed views. ``The layout of neuron similarities and important weights provides an alternative way of understanding the network layers and the links between layers. I can easily find important neurons in a specific layer and locate corresponding similar ones in the other model, so I don't need to check the activation maps of neurons back and forth between the source and the target models. While linking the IDs of important neurons and the neurons in the feature ranking list, I could identify whether a layer prefers domain-invariant or -specific features.''}

\revise{The experts also offered suggestions on how to improve the usability of our framework. E2 discussed the feasibility of supporting other network types, such as recurrent neural networks and auto-encoders. E1 suggested providing functions to support the comparison of multiple source and target models. ``We could investigate how the existing knowledge is preserved or ignored in different source-target pairs and various transfer learning settings simultaneously, which may help experts optimize the transfer strategies.''}
\section{Discussion and Conclusions}
\label{sec:discussion}

In this paper, we present a visual analytics framework for inspecting and exploring transfer learning processes. To provide a comprehensive analysis of knowledge transfer between deep neural networks, we have identified a set of analytical tasks to guide our framework design. In the visual analytics framework, the relationships between the models in two domains are presented with a multi-aspect design including statistical information, instance-level analysis, and comparative analysis of neural network components. Analysts can check model performances, inspect data distributions between classes, and diagnose knowledge transfer between neurons and weights. We demonstrate the usability of our visual analytics framework through case studies and expert interviews. An implementation of our framework is released on Github\footnote{https://github.com/VADERASU/visual-analytics-for-deep-transfer-learning}.


\vspace{0.8mm} \noindent \textbf{Scalability.} We discuss the issues of scalability from three aspects: automated algorithms, visual presentation, and task generalizability.

\vspace{0.6mm}
\noindent \textit{Automated Algorithms:} Due to the massive computational cost of deep neural networks, the running time for the extraction and comparison procedures can be time-consuming. All instances from both domains should be passed to the two models for retrieving the Layer Conductance values on all network layers. For the two datasets in the case study, it took about 1.5 and 3.5 hours respectively to compute the Layer Conductance values, the aggregations, and the importance rankings. The cost will increase significantly in complex network architectures such as ResNet and Inception. A feasible way to reduce the overhead is to run the extraction algorithm on a selected subset of layers.

\vspace{0.6mm}
\noindent \textit{Visual Presentation:} In our design, the visual clutter occurs in the t-SNE projection and the domain discriminability plot when large amounts of instances are in the selected classes. For the overplotting issue, a future solution to further scale our design is to adopt sampling and aggregation methods to remove unnecessary points~\cite{Mayorga2013,Chen2014Bluenoise,Liao2018}. In the network relation view, the neuron similarity matrices can be very large if there are too many important neurons. We can apply a filter to each matrix to limit the number of visible important neurons on demand.

\vspace{0.6mm}
\noindent \textit{Task Generalizability:} We use the basic fine-tuning method as a representative of the transfer learning methods for deep neural networks to illustrate our framework. Our framework supports various deep transfer learning approaches once they share the same protocol with the access of datasets as well as supporting layer attribution computation in the models from the two domains. Partial transfers can also be supported where only a subset of layers is shared between two models. To some extent, our framework can be further generalized as a one-to-one model comparison tool for general model selection.

\vspace{0.8mm} \noindent \textbf{\revise{Target Audience and Analysis Guidelines.}} The target audience for our framework are the machine learning practitioners and experts in different application domains where the transfer learning approaches are adopted in their daily workflow. \revise{For the practitioners, our framework can be used as a performance evaluation tool when target models are trained with the help of selected source domains. The insights gained from the visual exploration and inspection can further support the successive model selection. Similarly, the experts may also benefit from inspecting the knowledge transfer results when designing new transfer algorithms.} \revise{To better use our framework, we suggest starting the analysis by observing the accuracy chart and the confusion table to identify the desired class (which can be seen in the ``Analysis of the Model Statistics'' step in both cases), followed by inspecting data distributions from two domains in the instance view. The inspection of neuron similarities and importance rankings should then be considered after the class is activated in the network relation view, together with exploring the domain discriminability and the feature rankings. As demonstrated in ``In-depth Exploration of Models'' about distinguishing between bikes with different tire shapes, Section~\ref{sub:case1}, exploring the results in the network relation view and the feature view can benefit understanding the semantic differences for the same class between two domains.}


\vspace{0.8mm} \noindent \textbf{Future Work.} In the future, we expect to explore different neuron and weight extraction criteria to evaluate different neuron attribution methods in terms of revealing model similarities. \revise{To facilitate the analysis between multiple domains, we plan to support the analysis of more than one source domain and introduce better measurements on the transferability of different source domains.} Another promising extension is to enhance the usability of our framework in specific application scenarios such as object recognition and medical imaging.





\end{spacing}

\vspace{-0.7mm}
\acknowledgments{
This work was supported by the U.S. Department of Homeland Security under Grant Award 2017-ST-061-QA0001 and 17STQAC00001-03-03, and the National Science Foundation Program on Fairness in AI in collaboration with Amazon under award No. 1939725. The views and conclusions contained in this document are those of the authors and should not be interpreted as representing the official policies, either expressed or implied, of the U.S. Department of Homeland Security.
}

\bibliographystyle{abbrv-doi}

\bibliography{template}

\begin{thebibliography}{10}

\bibitem{tensorboard}
Tensorboard.
\newblock https://www.tensorflow.org/tensorboard.
\newblock Accessed: 2020-02-10.

\bibitem{abadi2016deep}
M.~Abadi, A.~Chu, I.~Goodfellow, H.~B. McMahan, I.~Mironov, K.~Talwar, and
  L.~Zhang.
\newblock Deep learning with differential privacy.
\newblock In {\em Proceedings of the ACM Conference on Computer and
  Communications Security}, pp. 308--318, 2016.

\bibitem{Ahn2019}
Y.~{Ahn} and Y.~{Lin}.
\newblock Fairsight: Visual analytics for fairness in decision making.
\newblock {\em IEEE Transactions on Visualization and Computer Graphics},
  26(1):1086--1095, 2020.

\bibitem{Alexander2016}
E.~Alexander and M.~Gleicher.
\newblock Task-driven comparison of topic models.
\newblock {\em IEEE Transactions on Visualization and Computer Graphics},
  22(1):320--329, 2016.

\bibitem{Alsallakh2017}
B.~Alsallakh, A.~Jourabloo, M.~Ye, X.~Liu, and L.~Ren.
\newblock Do convolutional neural networks learn class hierarchy?
\newblock {\em IEEE Transactions on Visualization and Computer Graphics},
  24(1):152--162, 2017.

\bibitem{Amershi}
S.~Amershi, M.~Chickering, S.~M. Drucker, B.~Lee, P.~Simard, and J.~Suh.
\newblock {ModelTracker}: Redesigning performance analysis tools for machine
  learning.
\newblock In {\em Proceedings of the ACM Conference on Human Factors in
  Computing Systems}, pp. 337--346, 2015.

\bibitem{antol2015vqa}
S.~Antol, A.~Agrawal, J.~Lu, M.~Mitchell, D.~Batra, C.~Lawrence~Zitnick, and
  D.~Parikh.
\newblock {VQA}: Visual question answering.
\newblock In {\em Proceedings of the IEEE International Conference on Computer
  Vision}, pp. 2425--2433, 2015.

\bibitem{ben2007analysis}
S.~Ben-David, J.~Blitzer, K.~Crammer, and F.~Pereira.
\newblock Analysis of representations for domain adaptation.
\newblock In {\em Advances in Neural Information Processing Systems}, pp.
  137--144, 2007.

\bibitem{Bertini2009}
E.~Bertini and D.~Lalanne.
\newblock Surveying the complementary role of automatic data analysis and
  visualization in knowledge discovery.
\newblock In {\em Proceedings of the ACM SIGKDD Workshop on Visual Analytics
  and Knowledge Discovery Integrating Automated Analysis with Interactive
  Exploration}, pp. 12--20, 2009.

\bibitem{Brehmer2013}
M.~Brehmer and T.~Munzner.
\newblock A multi-level typology of abstract visualization tasks.
\newblock {\em IEEE Transactions on Visualization and Computer Graphics},
  19(12):2376--2385, 2013.

\bibitem{Cabrera2019}
{\'{A}}.~A. Cabrera, W.~Epperson, F.~Hohman, M.~Kahng, J.~Morgenstern, and
  D.~H. Chau.
\newblock {FairVis}: Visual analytics for discovering intersectional bias in
  machine learning.
\newblock In {\em Proceedings of the IEEE Conference on Visual Analytics
  Science and Technology}, 2019.

\bibitem{Cao2011}
N.~Cao, D.~Gotz, J.~Sun, and H.~Qu.
\newblock {DICON}: Interactive visual analysis of multidimensional clusters.
\newblock {\em IEEE Transactions on Visualization and Computer Graphics},
  17(12):2581--2590, 2011.

\bibitem{Cashman2020}
D.~Cashman, A.~Perer, R.~Chang, and H.~Strobelt.
\newblock {Ablate, variate, and contemplate}: Visual analytics for discovering
  neural architectures.
\newblock {\em IEEE Transactions on Visualization and Computer Graphics},
  26(1):863--873, 2020.

\bibitem{Chen2014Bluenoise}
H.~{Chen}, W.~{Chen}, H.~{Mei}, Z.~{Liu}, K.~{Zhou}, W.~{Chen}, W.~{Gu}, and
  K.~{Ma}.
\newblock Visual abstraction and exploration of multi-class scatterplots.
\newblock {\em IEEE Transactions on Visualization and Computer Graphics},
  20(12):1683--1692, Dec 2014.

\bibitem{Chung2016}
S.~Chung, C.~Park, S.~Suh, K.~Kang, J.~Choo, and B.~C. Kwon.
\newblock {ReVACNN}: Steering convolutional neural network via real-time visual
  analytics.
\newblock In {\em Proceedings of KDD Workshop on Interactive Data Exploration
  and Analytics}, 2016.

\bibitem{dhamdhere2018important}
K.~Dhamdhere, M.~Sundararajan, and Q.~Yan.
\newblock How important is a neuron?
\newblock {\em arXiv preprint arXiv:1805.12233}, 2018.

\bibitem{Endert2017}
A.~Endert, W.~Ribarsky, C.~Turkay, B.~L.~W. Wong, I.~Nabney, I.~D. Blanco, and
  F.~Rossi.
\newblock The state of the art in integrating machine learning into visual
  analytics.
\newblock {\em Computer Graphics Forum}, 36(8):458--486, 2017.

\bibitem{Gleicher2017}
M.~Gleicher.
\newblock Considerations for visualizing comparison.
\newblock {\em IEEE Transactions on Visualization and Computer Graphics},
  24(1):413--423, 2017.

\bibitem{Goodfellow2015}
I.~Goodfellow, J.~Shlens, and C.~Szegedy.
\newblock Explaining and harnessing adversarial examples.
\newblock In {\em Proceedings of the International Conference on Learning
  Representations}, 2015.

\bibitem{Hohman2019b}
F.~Hohman, A.~Head, R.~Caruana, R.~DeLine, and S.~M. Drucker.
\newblock Gamut: A design probe to understand how data scientists understand
  machine learning models.
\newblock In {\em Proceedings of the ACM Conference on Human Factors in
  Computing Systems}, 2019.

\bibitem{Hohman2018}
F.~Hohman, M.~Kahng, R.~Pienta, and D.~H. Chau.
\newblock {Visual Analytics in Deep Learning: An Interrogative Survey for the
  Next Frontiers}.
\newblock {\em IEEE Transactions on Visualization and Computer Graphics},
  25(8):2674--2693, 2019.

\bibitem{Hohman2019a}
F.~Hohman, H.~Park, C.~Robinson, and D.~H. Chau.
\newblock Summit: Scaling deep learning interpretability by visualizing
  activation and attribution summarizations.
\newblock {\em IEEE Transactions on Visualization and Computer Graphics},
  26(1):1096--1106, 2020.

\bibitem{Huang2019}
Z.~Huang, Y.~Lu, E.~Mack, W.~Chen, and R.~Maciejewski.
\newblock Exploring the sensitivity of choropleths under attribute uncertainty.
\newblock {\em IEEE Transactions on Visualization and Computer Graphics},
  26(8):2576--2590, 2020.

\bibitem{Kahng2017a}
M.~Kahng, P.~Y. Andrews, A.~Kalro, and D.~H. Chau.
\newblock {ActiVis}: Visual exploration of industry-scale deep neural network
  models.
\newblock {\em IEEE Transactions on Visualization and Computer Graphics},
  24(1):88--97, 2018.

\bibitem{Kahng}
M.~Kahng, N.~Thorat, D.~Horng, P.~Chau, F.~B. Vi{\'e}gas, and M.~Wattenberg.
\newblock {GAN Lab}: Understanding complex deep generative models using
  interactive visual experimentation.
\newblock {\em IEEE Transactions on Visualization and Computer Graphics},
  25(1):310--320, 2019.

\bibitem{Krause2014}
J.~Krause, A.~Perer, and E.~Bertini.
\newblock {INFUSE}: Interactive feature selection for predictive modeling of
  high dimensional data.
\newblock {\em IEEE Transactions on Visualization and Computer Graphics},
  20(12):1614--1623, 2014.

\bibitem{Krause}
J.~Krause, A.~Perer, and K.~Ng.
\newblock Interacting with predictions: Visual inspection of black-box machine
  learning models.
\newblock In {\em Proceedings of the ACM Conference on Human Factors in
  Computing Systems}, pp. 5686--5697, 2016.

\bibitem{krizhevsky2012imagenet}
A.~Krizhevsky, I.~Sutskever, and G.~E. Hinton.
\newblock {ImageNet} classification with deep convolutional neural networks.
\newblock In {\em Proceedings of Advances in Neural Information Processing
  Systems}, pp. 1097--1105, 2012.

\bibitem{Kwon2018}
B.~C. Kwon, M.~J. Choi, J.~T. Kim, E.~Choi, Y.~B. Kim, S.~Kwon, J.~Sun, and
  J.~Choo.
\newblock {RetainVis}: Visual analytics with interpretable and interactive
  recurrent neural networks on electronic medical records.
\newblock {\em IEEE Transactions on Visualization and Computer Graphics},
  25(1):299--309, 2019.

\bibitem{Leino2018}
K.~Leino, S.~Sen, A.~Datta, M.~Fredrikson, and L.~Li.
\newblock Influence-directed explanations for deep convolutional networks.
\newblock In {\em Proceedings of IEEE International Test Conference}, 2018.

\bibitem{Li2020compare}
Y.~Li, T.~Fujiwara, Y.~K. Choi, K.~K. Kim, and K.-L. Ma.
\newblock A visual analytics system for multi-model comparison on clinical data
  predictions.
\newblock {\em Visual Informatics}, 4(2):122--131, 2020.

\bibitem{Liao2018}
H.~{Liao}, Y.~{Wu}, L.~{Chen}, and W.~{Chen}.
\newblock Cluster-based visual abstraction for multivariate scatterplots.
\newblock {\em IEEE Transactions on Visualization and Computer Graphics},
  24(9):2531--2545, 2018.

\bibitem{lim2019trust}
Z.~W. Lim, M.~L. Lee, W.~Hsu, and T.~Y. Wong.
\newblock Building trust in deep learning system towards automated disease
  detection.
\newblock In {\em Proceedings of the AAAI Conference on Artificial
  Intelligence}, vol.~33, pp. 9516--9521, 2019.

\bibitem{Liu2018a}
D.~Liu, W.~Cui, K.~Jin, Y.~Guo, and H.~Qu.
\newblock {DeepTracker}: Visualizing the training process of convolutional
  neural networks.
\newblock {\em ACM Transactions on Intelligent Systems and Technology},
  10(1):1--25, 2018.

\bibitem{Liu2018}
M.~Liu, S.~Liu, H.~Su, K.~Cao, and J.~Zhu.
\newblock Analyzing the noise robustness of deep neural networks.
\newblock In {\em Proceedings of the IEEE Conference on Visual Analytics
  Science and Technology}, vol.~2, pp. 60--71, 2018.

\bibitem{Liu2017a}
M.~Liu, J.~Shi, K.~Cao, J.~Zhu, and S.~Liu.
\newblock Analyzing the training processes of deep generative models.
\newblock {\em IEEE Transactions on Visualization and Computer Graphics},
  24(1):77--87, 2017.

\bibitem{Liu}
M.~Liu, J.~Shi, Z.~Li, C.~Li, J.~Zhu, and S.~Liu.
\newblock Towards better analysis of deep convolutional neural networks.
\newblock {\em IEEE Transactions on Visualization and Computer Graphics},
  23(1):91--100, 2016.

\bibitem{Liu2017}
S.~Liu, X.~Wang, M.~Liu, and J.~Zhu.
\newblock Towards better analysis of machine learning models: A visual
  analytics perspective.
\newblock {\em Visual Informatics}, 1(1):48--56, 2017.

\bibitem{Liu2017b}
S.~Liu, J.~Xiao, J.~Liu, X.~Wang, J.~Wu, and J.~Zhu.
\newblock Visual diagnosis of tree boosting methods.
\newblock {\em IEEE Transactions on Visualization and Computer Graphics},
  24(1):163--173, 2017.

\bibitem{Long2018}
M.~Long, Y.~Cao, Z.~Cao, J.~Wang, and M.~I. Jordan.
\newblock {Transferable Representation Learning with Deep Adaptation Networks}.
\newblock {\em IEEE Transactions on Pattern Analysis and Machine Intelligence},
  41(12):3071--3085, 2018.

\bibitem{lu2017recent}
J.~Lu, W.~Chen, Y.~Ma, J.~Ke, Z.~Li, F.~Zhang, and R.~Maciejewski.
\newblock Recent progress and trends in predictive visual analytics.
\newblock {\em Frontiers of Computer Science}, 11(2):192--207, 2017.

\bibitem{Lu2018}
J.~Lu, A.~Liu, F.~Dong, F.~Gu, J.~Gama, and G.~Zhang.
\newblock Learning under concept drift: A review.
\newblock {\em IEEE Transactions on Knowledge and Data Engineering},
  4347(c):1--18, 2018.

\bibitem{Lu2017}
Y.~Lu, R.~Garcia, B.~Hansen, M.~Gleicher, and R.~Maciejewski.
\newblock The state-of-the-art in predictive visual analytics.
\newblock {\em Computer Graphics Forum}, 36(3):539--562, 2017.

\bibitem{Ma2017}
Y.~Ma, W.~Chen, X.~Ma, J.~Xu, X.~Huang, R.~Maciejewski, and A.~K. Tung.
\newblock {EasySVM}: A visual analysis approach for open-box support vector
  machines.
\newblock {\em Computational Visual Media}, 3(2):161--175, 2017.

\bibitem{Ma2020}
Y.~Ma, T.~Xie, J.~Li, and R.~Maciejewski.
\newblock Explaining vulnerabilities to adversarial machine learning through
  visual analytics.
\newblock {\em IEEE Transactions on Visualization and Computer Graphics},
  26(1):1075--1085, 2020.

\bibitem{Ma2017transfer}
Y.~Ma, J.~Xu, X.~Wu, F.~Wang, and W.~Chen.
\newblock A visual analytical approach for transfer learning in classification.
\newblock {\em Information Sciences}, 390:54--69, 2017.

\bibitem{maaten2008visualizing}
L.~v.~d. Maaten and G.~Hinton.
\newblock Visualizing data using {t-SNE}.
\newblock {\em Journal of Machine Learning Research}, 9(Nov):2579--2605, 2008.

\bibitem{Mayorga2013}
A.~{Mayorga} and M.~{Gleicher}.
\newblock Splatterplots: Overcoming overdraw in scatter plots.
\newblock {\em IEEE Transactions on Visualization and Computer Graphics},
  19(9):1526--1538, 2013.

\bibitem{Ming2017}
Y.~Ming, Z.~Li, and Y.~Chen.
\newblock Understanding hidden memories of recurrent neural networks.
\newblock In {\em Proceedings of the IEEE Conference on Visual Analytics
  Science and Technology}, 2017.

\bibitem{Ming2018}
Y.~Ming, H.~Qu, and E.~Bertini.
\newblock {RuleMatrix: Visualizing and Understanding Classifiers with Rules}.
\newblock {\em IEEE Transactions on Visualization and Computer Graphics},
  25(1):342--352, 2018.

\bibitem{Muhlbacher2014}
T.~Muhlbacher, H.~Piringer, S.~Gratzl, M.~Sedlmair, and M.~Streit.
\newblock Opening the black box: Strategies for increased user involvement in
  existing algorithm implementations.
\newblock {\em IEEE Transactions on Visualization and Computer Graphics},
  20(12):1643--1652, 2014.

\bibitem{Murugesan}
S.~Murugesan, S.~Malik, F.~Du, E.~Koh, and T.~M. Lai.
\newblock {DeepCompare: Visual and Interactive Comparison of Deep Learning
  Model Performance}.
\newblock {\em IEEE Computer Graphics and Applications}, 39(5):47--59, 2019.

\bibitem{olah2017feature}
C.~Olah, A.~Mordvintsev, and L.~Schubert.
\newblock Feature visualization.
\newblock {\em Distill}, 2(11):e7, 2017.

\bibitem{Pan2010a}
S.~J. Pan and Q.~Yang.
\newblock A survey on transfer learning.
\newblock {\em IEEE Transactions on Knowledge and Data Engineering},
  22(10):1345--1359, 2010.

\bibitem{pan2008transfer}
S.~J. Pan, V.~W. Zheng, Q.~Yang, and D.~H. Hu.
\newblock Transfer learning for wifi-based indoor localization.
\newblock In {\em Proceedings of the Association for the Advancement of
  Artificial Intelligence Workshop}, vol.~6, 2008.

\bibitem{Pezzotti2017}
N.~Pezzotti, T.~H{\"{o}}llt, J.~{Van Gemert}, B.~P. Lelieveldt, E.~Eisemann,
  and A.~Vilanova.
\newblock {DeepEyes}: Progressive visual analytics for designing deep neural
  networks.
\newblock {\em IEEE Transactions on Visualization and Computer Graphics},
  24(1):98--108, 2018.

\bibitem{Pilh2012}
A.~Pilh, A.~Gribov, and A.~Unwin.
\newblock Comparing clusterings using bertin's idea.
\newblock {\em IEEE Transaction on Visualization and Computer Graphics},
  18(12):2506--2515, 2012.

\bibitem{raghu2019transfusion}
M.~Raghu, C.~Zhang, J.~Kleinberg, and S.~Bengio.
\newblock Transfusion: Understanding transfer learning for medical imaging.
\newblock In {\em Proceedings of Advances in Neural Information Processing
  Systems}, pp. 3342--3352, 2019.

\bibitem{Rauber2016}
P.~E. Rauber, S.~G. Fadel, A.~Falcao, A.~C. Telea, A.~X. Falc{\~{a}}o, and
  A.~C. Telea.
\newblock Visualizing the hidden activity of artificial neural networks.
\newblock {\em IEEE Transactions on Visualization and Computer Graphics},
  23(1):101--110, 2017.

\bibitem{Ren2016}
D.~Ren, S.~Amershi, B.~Lee, J.~Suh, J.~D. Williams, D.~Ren, J.~Suh, and
  S.~Amershi.
\newblock Squares: Supporting interactive performance analysis for multiclass
  classifiers.
\newblock {\em IEEE Transactions on Visualization and Computer Graphics},
  23(1):61--70, 2016.

\bibitem{saenko2010adapting}
K.~Saenko, B.~Kulis, M.~Fritz, and T.~Darrell.
\newblock Adapting visual category models to new domains.
\newblock In {\em Proceedings of the European Conference on Computer Vision},
  pp. 213--226. Springer, 2010.

\bibitem{Selvaraju2017}
R.~R. Selvaraju, M.~Cogswell, A.~Das, R.~Vedantam, D.~Parikh, and D.~Batra.
\newblock {Grad-CAM}: Visual explanations from deep networks via gradient-based
  localization.
\newblock In {\em Proceedings of the IEEE International Conference on Computer
  Vision}, pp. 618--626, 2017.

\bibitem{shrikumar2018computationally}
A.~Shrikumar, J.~Su, and A.~Kundaje.
\newblock Computationally efficient measures of internal neuron importance.
\newblock {\em arXiv preprint arXiv:1807.09946}, 2018.

\bibitem{Spinner2019}
T.~Spinner, U.~Schlegel, H.~Sch{\"{a}}fer, M.~El-Assady, H.~Schafer, and
  M.~El-Assady.
\newblock {explAIner}: A visual analytics framework for interactive and
  explainable machine learning.
\newblock {\em IEEE Transactions on Visualization and Computer Graphics},
  26(1):1064--1074, 2020.

\bibitem{Strobelt2019}
H.~Strobelt, S.~Gehrmann, M.~Behrisch, A.~Perer, H.~Pfister, and A.~M. Rush.
\newblock {Seq2seq-Vis}: A visual debugging tool for sequence-to-sequence
  models.
\newblock {\em IEEE Transactions on Visualization and Computer Graphics},
  25(1):353--363, 2019.

\bibitem{Strobelt2016a}
H.~Strobelt, S.~Gehrmann, B.~Huber, H.~Pfister, and A.~M. Rush.
\newblock Visual analysis of hidden state dynamics in recurrent neural
  networks.
\newblock {\em IEEE Transactions on Visualization and Computer Graphics},
  24(1):667--676, 2016.

\bibitem{Talbot2009}
J.~Talbot, B.~Lee, A.~Kapoor, and D.~S. Tan.
\newblock {EnsembleMatrix}: interactive visualization to support machine
  learning with multiple classifiers.
\newblock In {\em Proceedings of the ACM Conference on Human Factors in
  Computing Systems}, p. 1283, 2009.

\bibitem{Tan2018}
C.~Tan, F.~Sun, T.~Kong, W.~Zhang, C.~Yang, and C.~Liu.
\newblock A survey on deep transfer learning.
\newblock In {\em Proceedings of the International Conference on Artificial
  Neural Networks}, pp. 270--279. Springer, 2018.

\bibitem{Tzeng2005}
F.-Y. Tzeng and K.-L. Ma.
\newblock Opening the black box - {Data} driven visualization of neural
  networks.
\newblock In {\em Proceedings of the IEEE Visualization Conference}, pp.
  383--390, 2005.

\bibitem{VonLandesberger2011}
T.~von Landesberger, A.~Kuijper, T.~Schreck, J.~Kohlhammer, J.~J. van Wijk,
  J.~D. Fekete, and D.~W. Fellner.
\newblock Visual analysis of large graphs: State-of-the-art and future research
  challenges.
\newblock {\em Computer Graphics Forum}, 30(6):1719--1749, 2011.

\bibitem{Wang2018}
J.~Wang, L.~Gou, H.~W. Shen, and H.~Yang.
\newblock {DQNViz}: A visual analytics approach to understand deep
  {Q}-networks.
\newblock {\em IEEE Transactions on Visualization and Computer Graphics},
  25(1):288--298, 2019.

\bibitem{Wang2018a}
J.~Wang, L.~Gou, H.~Yang, and H.~W. Shen.
\newblock {GANViz}: A visual analytics approach to understand the adversarial
  game.
\newblock {\em IEEE Transactions on Visualization and Computer Graphics},
  24(6):1905--1917, 2018.

\bibitem{Wang2019deepvid}
J.~Wang, L.~Gou, W.~Zhang, H.~Yang, and H.~W. Shen.
\newblock {DeepVID}: Deep visual interpretation and diagnosis for image
  classifiers via knowledge distillation.
\newblock {\em IEEE Transactions on Visualization and Computer Graphics},
  25(6):2168--2180, 2019.

\bibitem{Wang2019atmseer}
Q.~Wang, Y.~Ming, Z.~Jin, Q.~Shen, D.~Liu, M.~J. Smith, K.~Veeramachaneni, and
  H.~Qu.
\newblock {AtmSeer}: Increasing transparency and controllability in automated
  machine learning.
\newblock In {\em Proceedings of the ACM Conference on Human Factors in
  Computing Systems}, 2019.

\bibitem{Wang2019genealogy}
Q.~Wang, J.~Yuan, S.~Chen, H.~Su, H.~Qu, and S.~Liu.
\newblock Visual genealogy of deep neural networks.
\newblock {\em IEEE Transactions on Visualization and Computer Graphics}, 2019.

\bibitem{Weiss2016}
K.~Weiss, T.~M. Khoshgoftaar, and D.~Wang.
\newblock {A survey of transfer learning}.
\newblock {\em Journal of Big Data}, 3(1), 2016.

\bibitem{Wexler2019}
J.~Wexler, M.~Pushkarna, T.~Bolukbasi, M.~Wattenberg, F.~Viegas, and J.~Wilson.
\newblock {The What-If Tool}: Interactive probing of machine learning models.
\newblock {\em IEEE Transactions on Visualization and Computer Graphics},
  26(1):56--65, 2020.

\bibitem{Wongsuphasawat2017}
K.~{Wongsuphasawat}, D.~{Smilkov}, J.~{Wexler}, J.~{Wilson}, D.~{Mané},
  D.~{Fritz}, D.~{Krishnan}, F.~B. {Viégas}, and M.~{Wattenberg}.
\newblock Visualizing dataflow graphs of deep learning models in tensorflow.
\newblock {\em IEEE Transactions on Visualization and Computer Graphics},
  24(1):1--12, 2018.

\bibitem{Yosinski2014}
J.~Yosinski, J.~Clune, Y.~Bengio, and H.~Lipson.
\newblock How transferable are features in deep neural networks?
\newblock In {\em Proceedings of Advances in Neural Information Processing
  Systems}, pp. 3320--3328, 2014.

\bibitem{Young2018nlp}
T.~{Young}, D.~{Hazarika}, S.~{Poria}, and E.~{Cambria}.
\newblock Recent trends in deep learning based natural language processing.
\newblock {\em IEEE Computational Intelligence Magazine}, 13(3):55--75, 2018.

\bibitem{Zeng2017}
H.~Zeng, H.~Haleem, X.~Plantaz, N.~Cao, and H.~Qu.
\newblock {CNNComparator}: Comparative analytics of convolutional neural
  networks.
\newblock In {\em Proceedings of Workshop on Visual Analytics for Deep Learning
  (VADL)}, 2017.

\bibitem{Zhang2018}
J.~Zhang, Y.~Wang, P.~Molino, L.~Li, and D.~S. Ebert.
\newblock Manifold: A model-agnostic framework for interpretation and diagnosis
  of machine learning models.
\newblock {\em IEEE Transactions on Visualization and Computer Graphics},
  25(1):364--373, 2018.

\bibitem{Zhang2017}
Y.~Zhang and R.~MacIejewski.
\newblock Quantifying the visual impact of classification boundaries in
  choropleth maps.
\newblock {\em IEEE Transactions on Visualization and Computer Graphics},
  23(1):371--380, 2017.

\bibitem{Zhao2019}
X.~Zhao, Y.~Wu, D.~L. Lee, and W.~Cui.
\newblock {iForest}: Interpreting random forests via visual analytics.
\newblock {\em IEEE Transactions on Visualization and Computer Graphics},
  25(1):407--416, 2019.

\end{thebibliography}
\end{document}